\documentclass[nofootinbib,superscriptaddress,aps,preprint,prc]{revtex4}
\usepackage[utf8]{inputenc}

\usepackage{amsmath}
\usepackage{mathrsfs}
\usepackage{amssymb}
\usepackage{bbm}
\usepackage{graphicx}
\usepackage{graphics}
\usepackage[small,compact]{titlesec}
\usepackage{xspace}
\usepackage{colordvi}
\usepackage{color}
\usepackage{units}
\usepackage{stackrel}
\usepackage{hyperref}

\usepackage{ulem}

\renewcommand{\emph}[1]{\textit{#1}}

\newcommand{\vect}[1]{\vec{\mathbf{#1}}}
\newcommand{\vectS}[1]{\vec{\boldsymbol{#1}}}

\newcommand{\EFT}{$\mathrm{EFT}_{\not{\pi}}$\xspace}

\newcommand{\comment}[1]{}
\newcommand{\srutTypeOne}[1]{\vrule width0pt height0pt depth #1\relax}
\newcommand{\skHe}{\mathrm{He}}
\newcommand{\skH}{\mathrm{H}}

\makeatletter

\newcommand{\Rmnum}[1]{\expandafter\@slowromancap\romannumeral #1@}
\newcommand{\vast}{\bBigg@{4}}
\newcommand{\Vast}{\bBigg@{5}}
\makeatother

\begin{document}

\title{${}^3\skHe$ and $pd$ Scattering to Next-to-Leading Order in Pionless Effective Field Theory}

\author{Jared Vanasse}
\email{jjv9@phy.duke.edu}
\affiliation{Department of Physics, Duke University, Durham, NC 27708, USA
}

\author{David A. Egolf}
\email{dae3@georgetown.edu}
\affiliation{Department of Physics, Georgetown University, Washington, DC 20057, USA
}
\author{John Kerin}
\email{jak247@georgetown.edu}
\affiliation{Department of Physics, Georgetown University, Washington, DC 20057, USA
}
\author{Sebastian K\"onig}
\email{koenig.389@physics.osu.edu}
\affiliation{Department of Physics, The Ohio State University, Columbus, Ohio 43210, USA}
\affiliation{Helmholtz-Institut f\"ur Strahlen- und Kernphysik (Theorie)\\
and Bethe Center for Theoretical Physics, Universit\"at Bonn, 53115 Bonn, Germany
}
\author{Roxanne P. Springer}
\email{rps@phy.duke.edu}
\affiliation{Department of Physics, Duke University, Durham, NC 27708, USA
}

\date{Februaray 25, 2014; revised: \today}

\begin{abstract}

We study the three-body systems of ${}^{3}\skHe$ and $pd$ scattering and demonstrate, both analytically and numerically, that a new $pd$ three-body force is needed at next-to-leading order in pionless effective field theory.  We also show that at leading order these observables require no new three-body force beyond what is necessary to describe $nd$ scattering.  We include electromagnetic effects by iterating only diagrams that involve a single photon exchange in the three-body sector. 

\end{abstract}


\maketitle

\section{Introduction}

There remain long-standing open questions in three-nucleon physics.  One example is the $A_y$ puzzle, where experimental evidence~\cite{Huttel:1983,PhysRevC.50.589,PhysRevC.52.1193} is not consistent with existing theoretical predictions~\cite{Kievsky:1996ca,Huber:1998hu,Entem:2001tj}.  For decades these systems were studied using various potential models~\cite{Gloeckle1996107}.  Now the technology of effective field theories (EFTs) has advanced to the point where we can address these issues using a systematic, QCD-symmetry based EFT to complement the predictions of potential models.  It is clear that the resolution of outstanding three-nucleon puzzles will require EFT calculations to high precision.  This paper is part of that effort.

For momenta below the threshold for producing dynamical pions ($p<\Lambda_{\not{\pi}}\sim m_{\pi}$), nuclear physics can be described by a Lagrangian that consists solely of contact interactions between and among nucleon fields and external currents.  This theory, pionless effective field theory (\EFT), has a simple and manifest power counting~\cite{vanKolck:1997ut,vanKolck:1998bw,Gegelia:1998gn,Kaplan:1998tg,Mehen:1998tp}.  In the two-nucleon sector \EFT has been used successfully to calculate nucleon--nucleon ($NN$) scattering~\cite{Kong:1999sf,Chen:1999tn,Ando:2004mm,Ando:2007fh}, electromagnetic form factors of the deuteron~\cite{Chen:1999bg}, and the neutron--proton capture process~\cite{Rupak:1999rk,Ando:2005cz}.  It has also been used to study $NN$ parity-violation~\cite{Savage:2000iv,Phillips:2008hn,Schindler:2009wd,Shin:2009hi} and neutrino--deuteron processes~\cite{Kong:2000px,Butler:2000zp,Ando:2008va,Chen:2012hm}.

In three-nucleon systems, \EFT has been used to calculate nucleon--deuteron ($Nd$) scattering~\cite{Bedaque:1999vb,Gabbiani:1999yv,Rupak:2001ci,Bedaque:2002yg,Griesshammer:2004pe,Konig:2011yq,Vanasse:2013sda}, ${}^{3}\skH$ and ${}^{3}\skHe$ bound-state properties~\cite{Ando:2010wq}, and parity-violation in $nd$ interactions~\cite{Griesshammer:2011md,Vanasse:2011nd}.  The case of $pd$ scattering in \EFT was originally investigated by Rupak and Kong~\cite{Rupak:2001ci}.  They treated Coulomb corrections perturbatively in $\alpha$, the fine structure constant, and developed a new power counting scheme in which the usual $Q$ counting was supplemented with an additional scale $p$, the external momentum.  They were able to calculate quartet S-wave $pd$ scattering to next-to-next-to leading order (NNLO) when certain diagrams were partially resummed~\cite{Gabbiani:1999yv,Griesshammer:2004pe} and found reasonable agreement with phase shift data.  However, their technique encountered numerical problems at center-of-mass (c.m.) momenta below 20 MeV.  Further, their calculation was not strictly perturbative in the \EFT power counting, but contained a subset of higher order terms.

The work presented here builds upon that of K\"onig and Hammer~\cite{Konig:2011yq} who, extending the previous work of Rupak and Kong, carried out calculations up to NNLO for both the quartet and doublet S-wave channels.  Using an optimized integration mesh they were able to obtain reasonable results down to a c.m. momentum of about 3 MeV.  However, again this calculation was not strictly perturbative in the \EFT power counting.  In addition, they assumed that up to next-to-leading order (NLO) the three-body forces from doublet S-wave $nd$ scattering were sufficient to obtain cutoff-independent results for $pd$ scattering.  Although their NLO phase shifts seem to have reasonable cutoff dependence, they did not go to large enough cutoffs to really test cutoff independence.  Indeed, we show here that at NLO, fixing a three-body force to only $nd$ physics yields $pd$ phase shifts and ${}^{3}\skHe$ binding energies that do not converge for large cutoffs.

The primary results of this paper are as follows:  We show analytically and numerically that at leading order (LO) no new three-body forces are needed in $pd$ scattering beyond those for $nd$ scattering.  However, we show that at NLO a new $pd$ three-body force is required to obtain cutoff-independent results for $pd$ scattering.  Without that force we see that for cutoffs much larger than 600 MeV there is significant cutoff variation in the NLO $pd$ phase shifts and NLO corrections to the ${}^{3}\skHe$ binding energy.  At NLO we fit this new three-body force to the ${}^{3}\skHe$ binding energy and show that we then obtain cutoff-independent results for the NLO $pd$ phase shifts.  We also calculate an analytical form for this three-body force and demonstrate agreement with the numerically calculated values.\footnote{After discovering the necessity for an $\alpha$-dependent three-body
force at NLO in the $ppn$ system, four of the authors became aware of
parallel work done by others in the field.  One of those (SK)
subsequently joined this paper as a fifth author; a part of his
analysis~\cite{Koenig:2013} is also presented here.  The authors then
became aware of previous discussions of the possibility of such a force
that took place between U. van Kolck and H.-W. Hammer, with additional
discussions among Hammer, D.R. Phillips, and SK.  Further work was then
carried out by SK, H.W. Grie{\ss}hammer, and Hammer.  A paper on this topic
is forthcoming~\cite{Konig:2014ufa}.}

\section{Effective Lagrangian}

The Lagrangian in the auxiliary field formalism up to NLO, including 
electromagnetic interactions and three-body forces, is given by
\begin{multline}
\label{eq:Lagrangian}
\mathcal{L}=\hat{N}^{\dagger}\left(iD_{0}+\frac{\vect{D}^2}{2M_{N}}\right)\hat{N}-\hat{t}_{i}^{\dagger}\left(iD_{0}+\frac{\vect{D}^{2}}{4M_{N}}-\Delta^{({}^3\!S_{1})}_{(-1)}-\Delta^{({}^3\!S_{1})}_{(0)}\right)\hat{t}_{i}+y_{t}\left[\hat{t}_{i}^{\dagger}\hat{N} ^{T}P_{i}\hat{N} +\mathrm{H.c.}\right]\\
-\hat{s}_{a}^{\dagger}\left(iD_{0}+\frac{\vect{D}^{2}}{4M_{N}}-\Delta^{({}^1\!S_{0})}_{(-1)}-\Delta^{({}^1\!S_{0})}_{(0)}\right)\hat{s}_{a}+y_{s}\left[\hat{s}_{a}^{\dagger}\hat{N}^{T}\bar{P}_{a}\hat{N}+\mathrm{H.c.}\right]+\mathcal{L}_{\mathrm{photon}}+\mathcal{L}_{3},
\end{multline}
where the deuteron field (spin-singlet dibaryon field) $\hat{t}_{i}$ ($\hat{s}_{a}$) is a spin-triplet iso-singlet (spin-singlet iso-triplet) combination of nucleons.  The projector $P_{i}=\frac{1}{\sqrt{8}}\sigma_{2}\sigma_{i}\tau_{2}$ ($\bar{P}_{a}=\frac{1}{\sqrt{8}}\tau_{2}\tau_{a}\sigma_{2}$) projects out the spin-triplet iso-singlet (spin-singlet iso-triplet) combination of nucleons.  The covariant derivative is defined by
\begin{equation}
D_{\mu}=\partial_{\mu}+ie\hat{A}_{\mu}\mathbf{Q},
\end{equation}
with the charge operator $\mathbf{Q}=1,(\mathbbm{1}+\tau_{3})/2,\mathbbm{1}+I_{3}$ for the $\mathbf{1},\mathbf{2}$, and $\mathbf{3}$ representations of SU(2) isospin, respectively ($I_{3}$ being the iso-triplet operator for isospin in the ``z''-direction).  The Lagrangian for pure photon contributions, $\mathcal{L}_{\mathrm{photon}}$, contains a kinetic and gauge fixing piece.  Since we need only static Coulomb exchange we keep only the temporal component of $\hat{A}_{\mu}$.  The propagator for the exchange of such potential photons is given by
\begin{equation}
i\Delta_{\mathrm{Coulomb}}(\vect{k})=\frac{i}{\vect{k}^{2}+\lambda^{2}},
\end{equation}
where $\lambda$ is a finite photon mass used to regulate both infrared and collinear divergences, and $\vect{k}$ is the photon three-momentum.  The results for zero photon mass are obtained by numerically extrapolating to the $\lambda=0$ limit.  

We do not need to include magnetic-moment interactions in our NLO
calculation.  Ref.~\cite{Rupak:1999rk} includes such effects at N$^2$LO in the spin-singlet $np$
channel, but generically, since compared to the leading Coulomb-photon
exchange they scale as ${Q^2 \over  M_N^2}$ and $M_N \gg \Lambda_{\not{\pi}} \sim  m_\pi$, such
effects are typically even smaller than N$^2$LO corrections (cf. the
counting of relativistic corrections in Refs.~\cite{Rupak:1999rk,Chen:1999tn}).

The Lagrangian for the three-body force is given by
\begin{align}
\label{eq:Lagrangian3body}
\mathcal{L}_{3}=&\frac{M_{N}H_{0}(\Lambda)}{3\Lambda^{2}}\left[y_{t}\hat{N}^{\dagger}(\vec{t}\cdot\vectS{\sigma})^{\dagger}-y_{s}\hat{N}^{\dagger}(\vec{s}\cdot\vectS{\tau})^{\dagger}\right]\left[y_{t}(\vec{t}\cdot\vectS{\sigma})\hat{N}-y_{s}(\vec{s}\cdot\vectS{\tau})\hat{N}\right]\\\nonumber
&+\frac{M_{N}H_{0}^{(\alpha)}(\Lambda)}{3\Lambda^{2}}\left[y_{t}\hat{N}^{\dagger}\mathbf{Q}(\vec{t}\cdot\vectS{\sigma})^{\dagger}-y_{s}\hat{N}^{\dagger}\mathbf{Q}\hat{s}_{3}^{\dagger}\tau^{3}-y_{s}\hat{N}^{\dagger}(\hat{s}_{1}\tau^{+})^{\dagger}\right]\times\\\nonumber
&\hspace{3cm}\left[y_{t}(\vec{t}\cdot{\vectS{\sigma}})\mathbf{Q}\hat{N}-y_{s}(\hat{s}_{3}\tau^{3})\mathbf{Q}\hat{N}-y_{s}(\hat{s}_{1}\tau^{+})\hat{N}\right],\\\nonumber
\end{align}
with $\tau^{+}=-(1/\sqrt{2})(\tau^{1}+i\tau^{2})$ and $H_{0}(\Lambda)$ and $H_{0}^{(\alpha)}(\Lambda)$ having explicit cutoff dependence to make the resulting physics cutoff-independent order by order in the \EFT expansion.  The expansion parameter of \EFT can be written as $\frac{Q}{\Lambda}\sim\gamma_t \rho_t$, which implies that the a priori estimate for the
uncertainty of a NLO calculation is $\mathcal{O}((\gamma_t\rho_t)^2)$, or roughly 17\%.

  $H_{0}(\Lambda)$ and $H_{0}^{(\alpha)}(\Lambda)$ are decomposed into LO, NLO, etc., pieces, yielding
\begin{equation}
H_{0}(\Lambda)=\underbrace{H_{0,0}(\Lambda)}_{\mathrm{LO}}+\underbrace{H_{0,1}(\Lambda)}_{\mathrm{NLO}}+\cdots
\end{equation}
and
\begin{equation}
H_{0}^{(\alpha)}(\Lambda)=\underbrace{H_{0,0}^{(\alpha)}(\Lambda)}_{\mathrm{LO}}+\underbrace{H_{0,1}^{(\alpha)}(\Lambda)}_{\mathrm{NLO}}+\cdots,
\end{equation}
so that $H_{0}(\Lambda)$ and $H_{0}^{(\alpha)}(\Lambda)$ need not be refit at each order.  At LO, $H_{0,0}(\Lambda)$ removes all cutoff dependence to order $(1/\Lambda)$, and $H^{(\alpha)}_{0,0}(\Lambda)=0$.  This is shown in Section~\ref{sec:LOAsymptotics}.  The NLO piece $H_{0,1}(\Lambda)$ removes linear and logarithmic divergences from the diagrams of $nd$ scattering at NLO.  A new feature that arises in the case of $pd$ scattering at NLO is the need for an $\alpha$-dependent three-body force $H_{0,1}^{(\alpha)}(\Lambda)$.  As shown in Section~ \ref{sec:NLO-3B-Phases}, including isospin breaking in the effective range for the $np$ and $pp$ singlet dibaryon propagators requires $H_{0,1}^{(\alpha)}(\Lambda)$ to remove both linear and logarithmic type divergences.  If isospin breaking effects in the effective range are ignored in $pd$ scattering, only logarithmic type divergences need to be removed by $H_{0,1}^{(\alpha)}(\Lambda)$.

\begin{figure}[hbt]
\includegraphics[width=140mm]{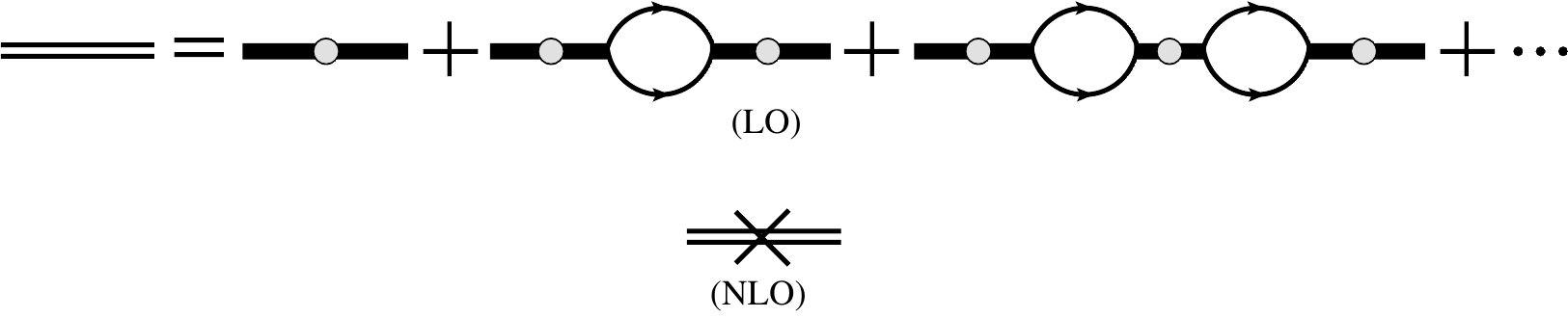}
\caption{\label{fig:LODeuteronPropagator} At LO the bare deuteron propagator $i/\Delta^{({}^{3}\!S_{1})}_{(-1)}$ is dressed by an infinite number of nucleon bubbles to give the LO dressed deuteron propagator.  At NLO the dressed deuteron propagator receives one effective range correction, which comes from the deuteron kinetic term and the NLO correction $\Delta^{({}^{3}\!S_{1})}_{(0)}$.}
\end{figure}

\begin{figure}[hbt]
\includegraphics[width=140mm]{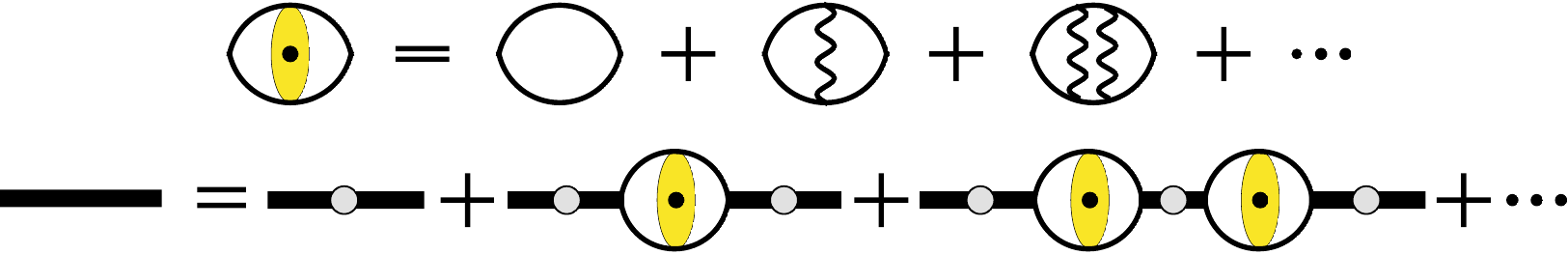}
\caption{\label{fig:LOppPropagator}(Color online) At LO the $pp$ nucleon bubble has an infinite series of ladder diagrams of Coulomb photon exchanges that can be resummed~\cite{Kong:1999sf,Holstein:2009zzb}.  The LO bare spin-singlet dibaryon propagator $i/\Delta^{({}^{1}\!S_{0})}_{(-1)}$ is dressed by an infinite number of $pp$ nucleon bubbles with photon ladder sums to give the LO dressed $pp$ dibaryon propagator.  At NLO the dressed $pp$ dibaryon propagator receives one effective range correction.
}
\end{figure}

In the Lagrangian of Eq.~(\ref{eq:Lagrangian}), the term $\Delta_{(0)}^{({}^{3}\!S_{1})}$ and the deuteron kinetic term are subleading compared to $\Delta_{(-1)}^{({}^{3}\!S_{1})}$. The bare deuteron propagator is given by $i/\Delta_{(-1)}^{({}^{3}\!S_{1})}$ and is dressed at LO by an infinite number of nucleon bubbles as in Fig.~\ref{fig:LODeuteronPropagator}.  
The sum of nucleon bubbles is a geometric series.  Unknown coefficients are fit to ensure that the deuteron pole is at the correct position.  At NLO the deuteron propagator gains a single insertion of the deuteron kinetic term and $\Delta_{(0)}^{({}^{3}\!S_{1})}$, as shown in Fig.~\ref{fig:LODeuteronPropagator}.  The $\Delta_{(-1)}^{({}^{3}\!S_{1})}$, $\Delta_{(0)}^{({}^{3}\!S_{1})}$, and $y_{t}$  coefficients are fit by ensuring that the deuteron pole is unchanged and that either (i) the deuteron pole has the correct residue, known as $Z$-parametrization; or (ii) the effective range expansion (ERE) about the deuteron pole is reproduced perturbatively, known as ERE parametrization~\cite{Phillips:1999hh,Griesshammer:2004pe}. For this paper we adopt the latter approach.  Details of this procedure and the resulting values of the coefficients have been discussed in Ref.~\cite{Griesshammer:2004pe}, so we merely quote the expression for the deuteron propagator to NLO in the ERE parametrization,
\begin{equation}
\label{eq:Dtprop}
iD_{t}(p_{0},\vect{p})=\frac{4\pi i}{M_{N}y_{t}^{2}}\frac{1}{\gamma_{t}-\sqrt{\frac{\vect{p}^{2}}{4}-M_{N}p_{0}-i\epsilon}}\left[\underbrace{\srutTypeOne{.5cm}1}_{\mathrm{LO}}-\underbrace{\frac{\rho_{t}}{2}\left(\sqrt{\frac{\vect{p}^{2}}{4}-M_{N}p_{0}-i\epsilon}+\gamma_{t}\right)}_{\mathrm{NLO}}\right].
\end{equation}
Here $\gamma_{t}=45.7025$~MeV is the deuteron binding momentum and $\rho_{t}=1.765$ fm is the effective range about the deuteron pole.  Analogously the spin-singlet dibaryon propagator to NLO is~\cite{Griesshammer:2004pe}
\begin{equation}
iD_{s}(p_{0},\vect{p})=\frac{4\pi i}{M_{N}y_{s}^{2}}\frac{1}{\gamma_{s}-\sqrt{\frac{\vect{p}^{2}}{4}-M_{N}p_{0}-i\epsilon}}\left[\underbrace{\srutTypeOne{.7cm}1}_{\mathrm{LO}}-\underbrace{\frac{\rho_{s}}{2}\frac{\frac{\vect{p}^{2}}{4}-M_{N}p_{0}}{\gamma_{s}-\sqrt{\frac{\vect{p}^{2}}{4}-M_{N}p_{0}-i\epsilon}}}_{\mathrm{NLO}}\right],
\end{equation}
where $\gamma_{s}=1/a_{s}$, $a_{s}=-23.714$~fm is the scattering length in the ${}^{1}\!S_{0}$ channel, and $\rho_{s}=2.73$~fm is the effective range in the ${}^{1}\!S_{0}$ channel in an expansion about zero momentum.  In the case where the spin-singlet dibaryon consists of two protons, Coulomb corrections must be included, as in Fig.~\ref{fig:LOppPropagator}.
All nucleon bubbles in this modified propagator have Coulomb-photon exchanges between the nucleons.  These exchanges can be resummed to all orders yielding the $pp$ dibaryon propagator to NLO~\cite{Kong:1999sf,Holstein:2009zzb}
\begin{equation}
iD_{pp}(p_{0},\vect{p})=\frac{4\pi i}{M_{N}y_{s}^{2}}\frac{1}{\frac{1}{a_{C}}+2\kappa H(\kappa/p')}\left[\underbrace{\srutTypeOne{.5cm}1}_{\mathrm{LO}}-\underbrace{\frac{r_{C}}{2}\frac{\frac{\vect{p}^{2}}{4}-M_{N}p_{0}}{\frac{1}{a_{
C}}+2\kappa H(\kappa/p')}}_{\mathrm{NLO}}\right],
\end{equation}
where
\begin{equation}
p'=i\sqrt{\frac{\vect{p}^{2}}{4}-M_{N}p_{0}-i\epsilon},\quad \kappa=\frac{\alpha M_{N}}{2},
\end{equation}
and
\begin{equation}
H(\eta)=\psi(i\eta)+\frac{1}{2i\eta}-\ln(i\eta).
\end{equation}
The function $\psi$ is the logarithmic derivative of the $\Gamma$ function.  The scattering length in the $pp$ channel is $a_{C}=-7.8063$~fm, and the effective range $r_{C}=2.794$~fm.

We label the propagators using the notation $D^{(n)}_{t}(p_{0},\vect{p})$ where 
$n=0,1$ refers to LO and NLO, respectively.  Thus $D^{(0)}_{t}(p_{0},\vect{p})$ is the LO deuteron propagator and $D^{(1)}_{t}(p_{0},\vect{p})$ contains only the NLO piece of the deuteron propagator, as labeled in Eq.~(\ref{eq:Dtprop}).  So $D_{t}(p_{0},\vect{p}) = D^{(0)}_{t}(p_{0},\vect{p})+D^{(1)}_{t}(p_{0},\vect{p})+ \cdots$.  Analogous labeling is used for the $np$ spin-singlet and $pp$ spin-singlet dibaryon propagators.

The deuteron wavefunction renormalization is given by the residue of the dressed deuteron propagator at the deuteron pole.  To simplify expressions for the amplitudes, we absorb into them a factor of $4/M_{N}$, which requires dividing the deuteron wavefunction renormalization by the same factor.  To NLO this yields
\begin{equation}
\label{eq:ZD}
Z_{D}=\frac{2\pi\gamma_{t}}{M_{N}y_{t}^{2}}\left[\underbrace{\srutTypeOne{.1cm}1}_{\mathrm{LO}}+\underbrace{\gamma_{t}\rho_{t}}_{\mathrm{NLO}}+\cdots\right],
\end{equation}
where $Z_{LO}=(2\pi\gamma_{t})/(M_{N}y_{t}^{2})$ and $Z_{NLO}=Z_{LO}\gamma_{t}\rho_{t}$, and $Z_{D}=Z_{LO}+Z_{NLO}+\cdots$.  Note that in the ERE parametrization the residue is approached perturbatively order by order and is not exact at NLO, unlike in the $Z$-parametrization.

\section{Coulomb Diagrams}

\begin{figure}[hbt]
\includegraphics[width=140mm]{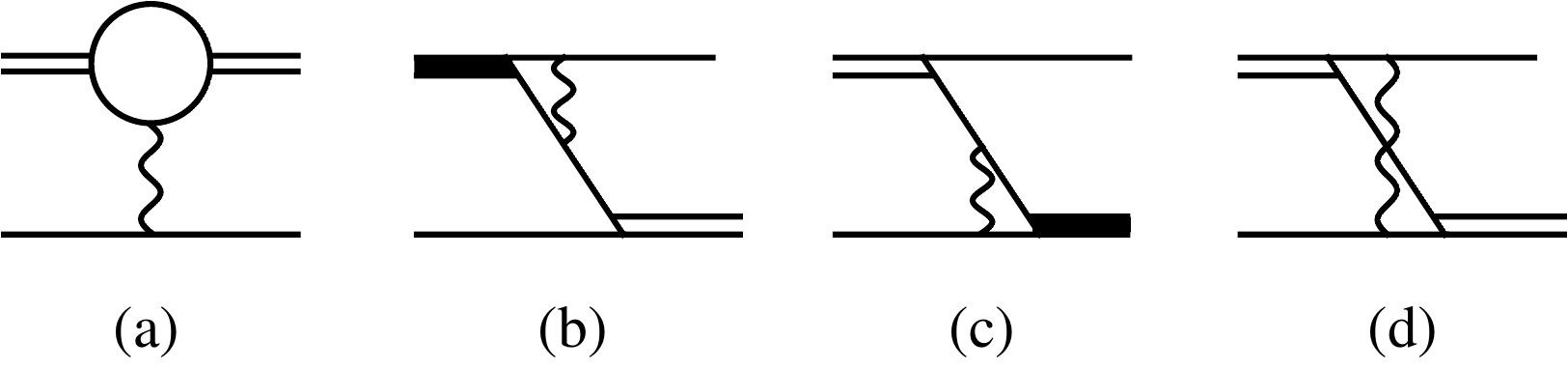}
\caption{\label{fig:CoulombLODiagrams} Coulomb corrections at LO.  Single lines are nucleon propagators, wavy lines are Coulomb photon propagators, double lines are either $np$ spin-singlet or spin-triplet dibyarons, and the thick solid lines are $pp$ dibaryons.}
\end{figure}

For this calculation we will use the power counting scheme for $pd$ scattering introduced by Rupak and Kong~\cite{Rupak:2001ci}.  In their scheme the usual $Q\sim\gamma_{t}$ counting is supplemented by a new scale for the external momentum, $p$.  Coulomb contributions scale as $\alpha M_{N}/p$. For low momentum transfers these will dominate over the scale $Q$ from strong physics.  For momenta $p\geq Q$ the usual $Q$ counting is recaptured.  The loop integration measure is $q^{0}q^{3}$.  In this power counting scheme, $q^0 \sim Q^2/M_{N}$, and $q$ either scales as $Q$ or $p$, depending upon whether the diagram is dominated by the external momentum $p$ or the binding momentum $\gamma_{t} \sim Q$.  In the integrand, dressed dibaryon propagators scale as $Q/q^{2}$ and photon propagators as $1/q^{2}$.  Nucleon lines scale as $M_{N}/Q^{2}$.
Using this power counting scheme the diagrams in Fig.~\ref{fig:CoulombLODiagrams} contribute at LO.  
With $y_{t}^{2}\sim y_{s}^{2}\sim \Lambda_{\not{\pi}}/M_{N}^{2}$~\cite{Gabbiani:1999yv}, where $\Lambda_{\not{\pi}}\sim m_{\pi}$ is the cutoff of \EFT, diagram~(a) scales as $\alpha \Lambda_{\not{\pi}}/p^{2}Q$ and diagrams~(b), (c), and~(d) all scale as $\alpha \Lambda_{\not{\pi}}/Q^{3}$.  At low momentum, diagram~(a) has an infrared divergence since it scales as $1/p^{2}$.  However, this divergence will be regulated by a finite photon mass.  The remaining diagrams (b)-(d) are infrared finite.  Both ${}^{3}\skHe$ and ${}^{3}\skH$ have a bound state momentum of roughly 75~MeV.  For this momentum, diagrams (a)-(d) are equally important; numerically we show that all diagrams are equally important in predicting the correct ${}^{3}\skHe$ bound state energy.  Calculations for $pd$ scattering have been carried out in Ref.~\cite{Konig:2011yq} for both the quartet and doublet S-wave channel Coulomb-subtracted phase shifts.  In that calculation, as in the earlier one by Rupak and Kong~\cite{Rupak:2001ci}, diagram~(d) 
is dropped because it is a 7\% effect at zero momentum, and diagrams~(b) and (c), contributing each at the 15\% level, are also dropped.  (In addition diagram~(a) is approximated using an on-shell approximation in which the dynamics from the nucleon bubble are neglected.)

These approximations yield good agreement with available phase shift data.  However, it is not legitimate to apply them in the bound-state regime, and so we do not use them in this paper.  Rather, we include all diagrams shown in Fig.~\ref{fig:CoulombLODiagrams} and keep the full dynamical expression for diagram~(a).  Projecting diagram~(a) onto the S-wave channel yields the analytical form
\begin{equation}
\label{eq:Bubble}
B(q,p,E)=\frac{4\alpha M_{N}}{qp}F_{1}\left[\lambda,2\sqrt{\frac{3}{4}q^{2}-M_{N}E-i\epsilon}+2\sqrt{\frac{3}{4}p^{2}-M_{N}E-i\epsilon},\vect{q}-\vect{p}\right],
\end{equation}
where $\vect{q}$ is the relative incoming three-momentum, $q$ its magnitude, $\vect{p}$ is the relative outgoing three-momentum, $p$ its magnitude, and $E$ the total energy of the system.
The function $F_{1}[a,b,\vect{c}+\vect{d}]$ for $\mathrm{Re}(b)>\mathrm{Re}(a)$ is defined as
\begin{align}
F_{1}[a,b,\vect{c}+\vect{d}]=&-\frac{1}{4a}\left\{\ln(z^{2}+a^{2})\ln\left(\frac{b-a}{b+a}\right)-\mathrm{Li}_{2}\left(-i\frac{z-ia}{b-a}\right)\right.\\\nonumber
&\hspace{2cm}\left.+\mathrm{Li}_{2}\left(i\frac{z-ia}{a+b}\right)-\mathrm{Li}_{2}\left(i\frac{z+ia}{b-a}\right)+\mathrm{Li}_{2}\left(-i\frac{z+ia}{a+b}\right)\right\}\stackbin[|c-d|]{c+d}{\Big{|}},
\end{align}
and for $\mathrm{Re}(a)>\mathrm{Re}(b)$ as
\begin{align}
F_{1}[a,b,\vect{c}+\vect{d}]=&\frac{1}{a}\tan^{-1}\left(\frac{z}{b}\right)\tan^{-1}\left(\frac{z}{a}\right)\\\nonumber
&+\frac{1}{4a}\left\{\ln(z^{2}+b^{2})\ln\left(\frac{a-b}{b+a}\right)-\mathrm{Li}_{2}\left(-i\frac{z-ib}{a-b}\right)\right.\\\nonumber
&\hspace{2cm}\left.+\mathrm{Li}_{2}\left(i\frac{z-ib}{a+b}\right)-\mathrm{Li}_{2}\left(i\frac{z+ib}{a-b}\right)+\mathrm{Li}_{2}\left(-i\frac{z+ib}{a+b}\right)\right\}\stackbin[|c-d|]{c+d}{\Big{|}},
\end{align}
where the bar notation is defined as
\begin{equation}
f(z)\stackbin[|c-d|]{c+d}{\Big{|}}=f(c+d)-f(|c-d|).
\end{equation}
A similar calculation for diagram~(b) yields
\begin{equation}
\label{eq:Vertex}
V_{1}(q,p,E)=\frac{4\alpha M_{N}}{qp}F_{1}\left[2\sqrt{\frac{3}{4}q^{2}-M_{N}E-i\epsilon},2\sqrt{\frac{3}{4}q^{2}-M_{N}E-i\epsilon}+2\lambda,\vect{q}+2\vect{p}\right].
\end{equation}
The S-wave projection of diagram~(c), $V_{2}(q,p,E)$, is related to that of diagram~(b) by time reversal symmetry:
\begin{equation}
\label{eq:v2}
V_{2}(q,p,E)=V_{1}(p,q,E).
\end{equation}
Diagram~(d) is more challenging.  In principle it can be solved and projected out in the S-wave channel exactly~\cite{Bedaque:1999vb}.  However, the resulting form is too lengthy and cumbersome for practical computation.  Instead, for $\lambda\ll\gamma_{t}$ we expand diagram~(d) in powers of $\lambda$~\cite{Hoferichter-BoxDiag:2010}.  Keeping all terms linear in $\lambda$ yields
\begin{align}
\label{eq:Cross}
&C(q,p,E)=-2\alpha M_{N}\times\\\nonumber
&\left(\frac{2}{qp}F_{1}\left[\sqrt{2M_{N}E-3q^{2}-3p^{2}+i\epsilon},2\sqrt{\frac{3}{4}q^{2}-M_{N}E-i\epsilon}+2\sqrt{\frac{3}{4}p^{2}-M_{N}E-i\epsilon},\vect{q}-\vect{p}\right]\right.\\\nonumber
&+\lambda\frac{1}{(p^{2}+q^{2}-M_{N}E-i\epsilon)^{2}-p^{2}q^{2}}+\mathcal{O}(\lambda^{2})+\cdots\Bigg{)}.
\end{align}
for the S-wave projected version of diagram~(d).

\section{Leading-Order Scattering Amplitude}

\begin{figure}[hbt!]
\includegraphics[width=150mm]{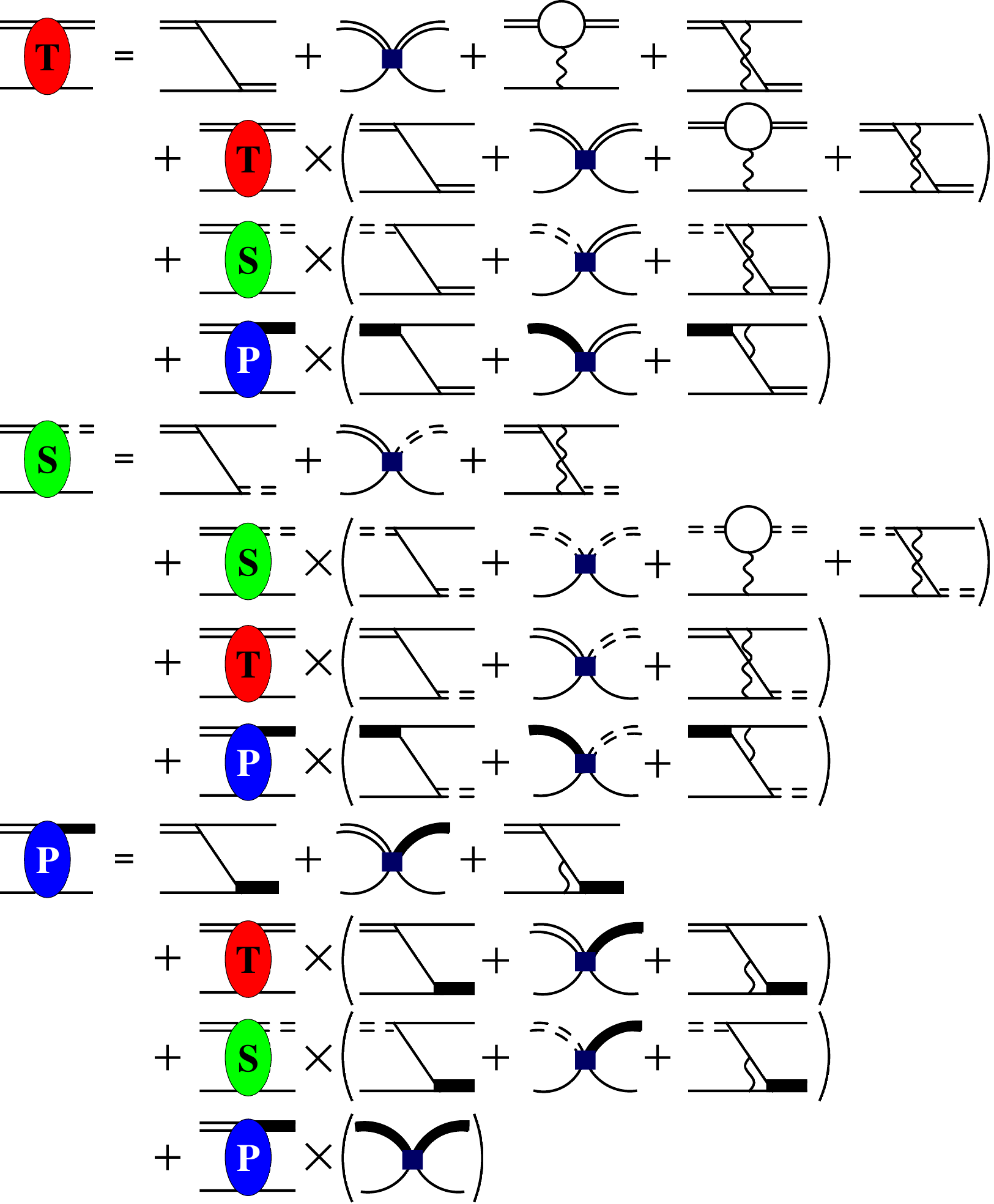}
\caption{\label{fig:LOpdScattering}(Color online) Coupled integral equations for LO doublet $pd$ scattering. The single line represents a nucleon, the double line a dressed deuteron propagator, the double-dashed line a dressed $np$ spin-singlet dibaryon propagator, and the thick solid line a $pp$ spin-singlet dibaryon propagator. The filled square is the three-body force $H_{0,0}(\Lambda)$.  The wavy lines are Coulomb photon exchanges.}
\end{figure}

The LO $pd$ scattering amplitude is found by solving the set of coupled integral equations shown in Fig.~\ref{fig:LOpdScattering}.
%
%
The ovals with a capital ``T'' represent the $t_{0,Nt\to Nt}(k,p,E)$ amplitude, ``S'' the amplitude $t_{0,Nt \to Ns}(k,p,E)$, and ``P'' the amplitude $t_{0,Nt\to Npp}(k,p,E)$, where the subscript 0 labels LO, $t$ labels the spin-triplet ``deuteron," $s$ labels the spin-singlet $np$ dibaryon, and $pp$ labels the spin-singlet $pp$ dibaryon.  The relative incoming momentum is $\vect{k}$ and the relative outgoing momentum is $\vect{p}$; see Fig. 5 in Ref.~\cite{Bedaque:1999vb} for momentum assignments.  Projecting the diagrams in Fig.~\ref{fig:LOpdScattering} onto the doublet S-wave channel, the scattering amplitude in cluster-configuration space~\cite{Griesshammer:2004pe} at LO is
\begin{equation}
\label{eq:LOScatt}
\mathbf{t}_{0}(k,p,E)=\mathbf{B}_{0}(k,p,E)+\mathbf{K}_{0}(q,p,E)\otimes\mathbf{t}_{0}(k,q,E).
\end{equation}
The subscript $0$ refers to LO and the bold script indicates that this is a matrix equation in cluster configuration space.  The amplitude $\mathbf{t}_{0}(k,p,E)$ is a three-vector defined by
\begin{align}
\mathbf{t}_{0}(k,p,E)=\left(\begin{array}{c}
t_{0,Nt\to Nt}(k,p,E)\\ 
t_{0,Nt\to Ns}(k,p,E)\\ 
t_{0,Nt\to Npp}(k,p,E)
\end{array}\right),
\end{align}
with $t_{0,Nt\to Nt}(k,p)$ the amplitude for $pd$ scattering, $t_{0,Nt\to Ns}(k,p)$ the amplitude for $pd$ going to a proton and an $np$ spin-singlet dibaryon, and $t_{0,Nt \to Npp}(k,p)$ the amplitude for $pd$ going to a neutron and a $pp$ spin-singlet dibaryon.  The ``$\otimes$'' operation is defined as
\begin{equation}
A(q)\otimes B(q)=\frac{2}{\pi}\int_{0}^{\Lambda}dqq^{2}A(q)B(q).
\end{equation}
The kernel and inhomogeneous terms are each decomposed into three pieces:
\begin{equation}
\mathbf{B}_{0}(k,p,E)=\mathbf{B}_{0}^{(S)}(k,p,E)+\mathbf{B}_{0}^{(SC)}(k,p,E)+\mathbf{B}_{0}^{(C)}(k,p,E),  
\end{equation}
and
\begin{equation}
\mathbf{K}_{0}(q,p,E)=\mathbf{K}_{0}^{(S)}(q,p,E)+\mathbf{K}_{0}^{(SC)}(q,p,E)+\mathbf{K}_{0}^{(C)}(q,p,E).
\end{equation}

The superscript $(S)$ refers to all contributions with only strong interactions, $(SC)$ to contributions that mix strong and Coulomb interactions, and $(C)$ to contributions containing only Coulomb interactions between the proton and remaining dibaryon field.  The inhomogeneous term 
\begin{align}
\label{eq:inhomDoubletS}
&\mathbf{B}^{(S)}_{0}(k,p,E)=\left(\begin{array}{c}
2y_{t}^{2}\left[\frac{1}{pk}Q_{0}\left(\frac{p^{2}+k^{2}-M_{N}E-i\epsilon}{pk}\right)+\frac{2H_{0,0}(\Lambda)}{\Lambda^{2}}\right]\\
2y_{t}y_{s}\left[\frac{1}{pk}Q_{0}\left(\frac{p^{2}+k^{2}-M_{N}E-i\epsilon}{pk}\right)+\frac{2H_{0,0}(\Lambda)}{3\Lambda^{2}}\right]\\
2y_{t}y_{s}\left[\frac{2}{pk}Q_{0}\left(\frac{p^{2}+k^{2}-M_{N}E-i\epsilon}{pk}\right)+\frac{4H_{0,0}(\Lambda)}{3\Lambda^{2}}\right]
\end{array}\right).\\\nonumber
\end{align}
The kernel matrix $\mathbf{K}_{0}^{(S)}(k,q,E)$ is defined by
\begin{align}
\label{eq:homDoubletS}
\mathbf{K}^{(S)}_{0}(q,p,E)=&\frac{M_{N}}{8\pi}\frac{1}{qp}Q_{0}\left(\frac{q^{2}+p^{2}-M_{N}E-i\epsilon}{qp}\right)\left(\begin{array}{ccc}
-y_{t}^{2} & -3y_{t}y_{s} & -3 y_{t}y_{s} \\
-y_{s}y_{t} & y_{s}^{2} &-y_{s}^{2} \\
-2y_{s}y_{t} & -2y_{s}^{2} & 0
\end{array}\right)\\\nonumber
&\times\mathbf{D}^{(0)}\left(E-\frac{\vect{q}^{2}}{2M_{N}},\vect{q}\right)\\\nonumber
&+\frac{M_{N}}{8\pi}\frac{2H_{0,0}(\Lambda)}{\Lambda^{2}}\left(\begin{array}{ccc}
-y_{t}^{2} & -y_{t}y_{s} & -y_{t}y_{s}\\
-\frac{1}{3}y_{s}y_{t} &-\frac{1}{3}y_{s}^{2} & -\frac{1}{3}y_{s}^{2}\\
-\frac{2}{3}y_{s}y_{t} & -\frac{2}{3}y_{s}^{2} & -\frac{2}{3}y_{s}^{2}
\end{array}\right)\mathbf{D}^{(0)}\left(E-\frac{\vect{q}^{2}}{2M_{N}},\vect{q}\right),
\end{align}
where $\mathbf{D}^{(0)}(E,\vect{q})$ is a matrix of dibaryon propagators given by
\begin{equation}
\label{eq:DibMatrix}
\mathbf{D}^{(0)}(E,\vect{q})=
\left(
\begin{array}{ccc}
D^{(0)}_{t}(E,\vect{q}) & 0 & 0\\
0 & D^{(0)}_{s}(E,\vect{q}) & 0\\
0 & 0 &  D^{(0)}_{pp}(E,\vect{q})
\end{array}\right).
\end{equation}
The function $Q_{0}(a)$ is a Legendre function of the second kind,
\begin{equation}
Q_{0}(a)=\frac{1}{2}\ln\left(\frac{a+1}{a-1}\right).
\end{equation}
The inhomogeneous term 
\begin{align}
\label{eq:inhomDoubletSC}
\mathbf{B}^{(SC)}_{0}(k,p,E)=\left(\begin{array}{c}
-y_{t}^{2}C(k,p,E)\\
-y_{t}y_{s}C(k,p,E)\\
-2y_{t}y_{s}V_{2}(k,p,E)
\end{array}\right),
\end{align}
and the kernel
\begin{align}
\label{eq:homDoubletSC}
&\mathbf{K}^{(SC)}_{0}(q,p,E)=\frac{M_{N}}{16\pi}\\\nonumber
&\times\left(\begin{array}{ccc}
y_{t}^{2}C(q,p,E) & 3y_{t}y_{s}C(q,p,E) & 3 y_{t}y_{s}V_{1}(q,p,E) \\
y_{s}y_{t} C(q,p,E) & -y_{s}^{2}C(q,p,E) &y_{s}^{2} V_{1}(q,p,E) \\
2y_{s}y_{t} V_{2}(q,p,E) & 2y_{s}^{2} V_{2}(q,p,E) & 0
\end{array}\right)\mathbf{D}^{(0)}\left(E-\frac{\vect{q}^{2}}{2M_{N}},\vect{q}\right).
\end{align}
Finally, the inhomogeneous term 
\begin{align}
\label{eq:inhomDoubletC}
\mathbf{B}^{(C)}_{0}(k,p,E)=\left(\begin{array}{c}
-y_{t}^{2}B(k,p,E)\\
0\\
0
\end{array}\right),
\end{align}
and the corresponding kernel is
\begin{align}
\label{eq:homDoubletC}
\mathbf{K}^{(C)}_{0}(q,p,E)=\frac{M_{N}}{16\pi}\left(\begin{array}{ccc}
y_{t}^{2}B(q,p,E) & 0 & 0 \\
0 & y_{s}^{2} B(q,p,E)  &0 \\
0 & 0 & 0
\end{array}\right)\mathbf{D}^{(0)}\left(E-\frac{\vect{q}^{2}}{2M_{N}},\vect{q}\right).
\end{align}

\section{Next-to-Leading-Order Scattering Amplitude}

\begin{figure}[hbt]
\includegraphics[width=160mm]{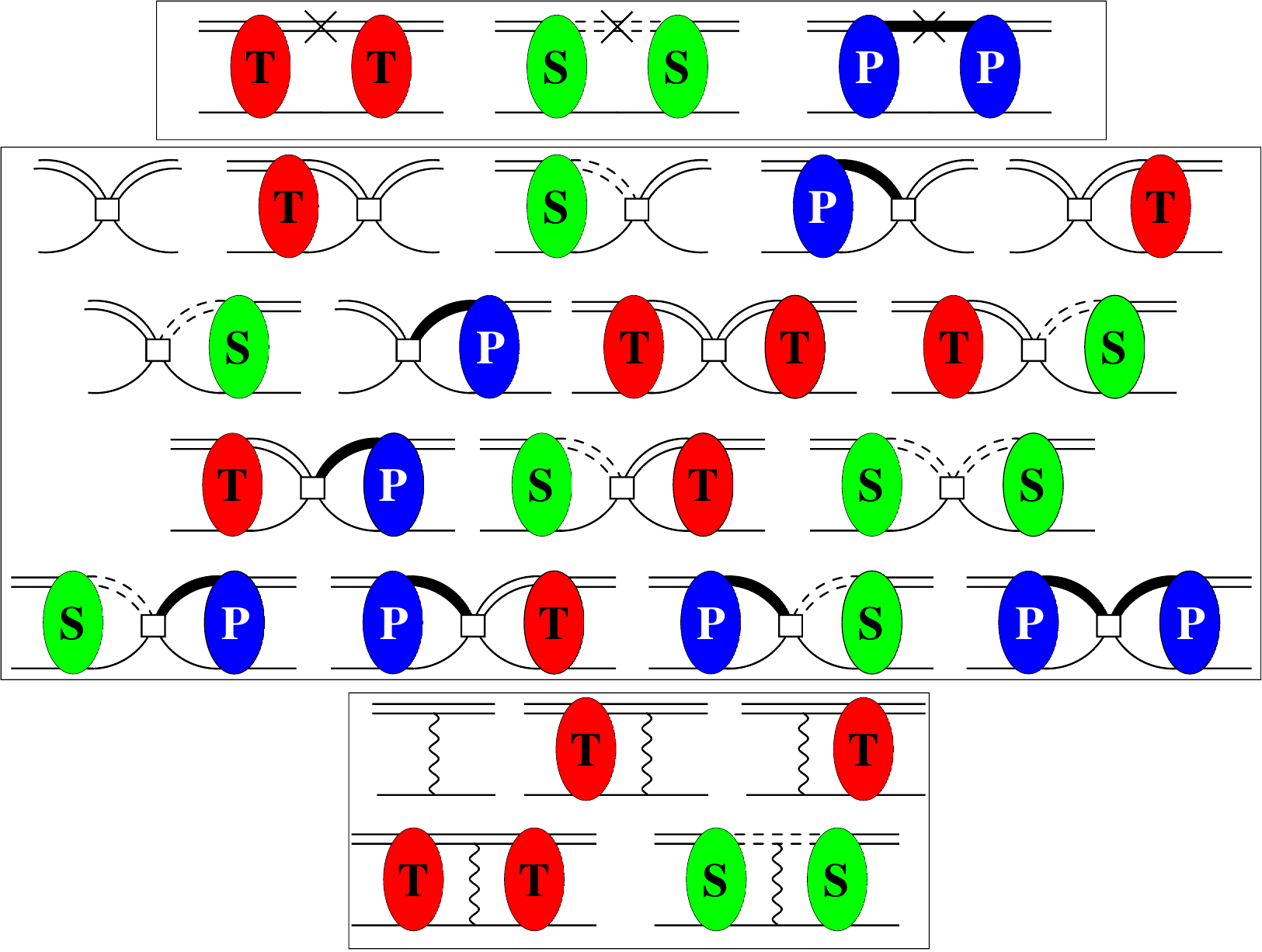}
\caption{\label{fig:NLOpdScattering}(Color online) NLO diagrams for $pd$ scattering.  The cross in the first boxed set of diagrams denotes a single insertion of an effective range correction.  All three-body force terms contain only the NLO and NLO-$\alpha$ three-body force and are depicted by blank squares. For other notation see Fig.~\ref{fig:LOpdScattering}. }
\end{figure}

The NLO $pd$ scattering amplitude is given by the sum of diagrams shown in Fig.~\ref{fig:NLOpdScattering}.
The letters in the ovals denote the same scattering amplitudes as in the LO case.  For the first boxed set of diagrams, labeled $t_{1}^{(ER)}(k,p,E)$, the cross represents an effective range insertion and makes the propagator between the two scattering amplitudes the NLO correction to the dibaryon propagator.  The diagrams in the second boxed set, labeled $t_{1}^{(3B)}(k,p,E)$, contain NLO three-body force terms represented by blank squares.  This blank square contains contributions from both $H_{0,1}(\Lambda)$ and $H_{0,1}^{\alpha}(\Lambda)$.  The final set of boxed diagrams, labeled $t_{1}^{(DK)}(k,p,E)$, contains the NLO Coulomb corrections that come from gauging the dibaryon kinetic term.  Splitting up the NLO $pd$ scattering amplitude into these three contributions yields
\begin{equation}
t_{1,Nt \to Nt}(k,p,E)=t_{1}^{(ER)}(k,p,E)+t_{1}^{(3B)}(k,p,E)+t_{1}^{(DK)}(k,p,E),
\end{equation}
where 
\begin{align}
\label{eq:NLOt1}
t_{1}^{(ER)}(k,p,E)=&\frac{\rho_{t}}{4\pi}\int_{0}^{\Lambda}\!\!dq q^{2}\left(t_{0,Nt\to Nt}(k,q)\right)^{2}\frac{\sqrt{\frac{3}{4}q^{2}-M_{N}E-i\epsilon}+\gamma_{t}}{\sqrt{\frac{3}{4}q^{2}-M_{N}E-i\epsilon}-\gamma_{t}}\\\nonumber
&+\frac{3\rho_{s}}{4\pi}\int_{0}^{\Lambda}\!\!dq q^{2}\left(t_{0,Nt\to Ns}(k,q)\right)^{2}\frac{\frac{3}{4}q^{2}-M_{N}E}{\left(\sqrt{\frac{3}{4}q^{2}-M_{N}E-i\epsilon}-\gamma_{s}\right)^{2}}\\\nonumber
&+\frac{3r_{C}}{8\pi}\int_{0}^{\Lambda}\!\!dq q^{2}\left(t_{0,Nt\to pp}(k,q)\right)^{2}\frac{\frac{3}{4}q^{2}-M_{N}E}{\left(2\kappa H\left(\frac{\kappa}{\sqrt{\frac{3}{4}q^{2}-M_{N}E-i\epsilon}}\right)+\frac{1}{a_{C}}\right)^{2}},
\end{align}
\begin{align}
\label{eq:NLOt2}
t_{1}^{(3B)}(k,p,E)&=\frac{4(H_{0,1}(\Lambda)+H_{0,1}^{(\alpha)}(\Lambda))}{\Lambda^{2}}\\\nonumber
&\times\vast[1+\frac{1}{2\pi}\int_{0}^{\Lambda}\!\!dqq^{2}t_{0,Nt\to Nt}(k,q)\frac{1}{\sqrt{\frac{3}{4}q^{2}-M_{N}E-i\epsilon}-\gamma_{t}}.\\\nonumber
&\hspace{2em}+\frac{1}{2\pi}\int_{0}^{\Lambda}\!\!dqq^{2}t_{0,Nt\to Ns}(k,q)\frac{1}{\sqrt{\frac{3}{4}q^{2}-M_{N}E-i\epsilon}-\gamma_{s}}\\\nonumber
&\hspace{2em}+\frac{1}{2\pi}\int_{0}^{\Lambda}\!\!dqq^{2}t_{0,Nt\to Npp}(k,q)\frac{1}{-\frac{1}{a_{C}}-2\kappa
H\left(\frac{\kappa}{\sqrt{\frac{3}{4}q^{2}-M_{N}E-i\epsilon}}\right)}\vast]^{2},
\end{align}
and
\begin{align}
\label{eq:NLOt3}
&t_{1}^{(DK)}(k,p,E)=-\frac{\alpha M_{N}\rho_{t}}{k^{2}}Q_{0}\left(\frac{2k^{2}+\lambda^{2}}{-2k^{2}}\right)\\\nonumber
&-\frac{\alpha M_{N}\rho_{t}}{\pi}\int_{0}^{\Lambda}\!\! dqq^{2}t_{0,Nt\to Nt}(k,q)\frac{1}{\sqrt{\frac{3}{4}q^{2}-M_{N}E-i\epsilon}-\gamma_{t}}\frac{1}{qk}Q_{0}\left(\frac{k^{2}+q^{2}+\lambda^{2}}{-2qk}\right)\\\nonumber
&-\frac{\rho_{t}\alpha M_{N}}{4\pi^{2}}\int_{0}^{\Lambda}\!\!dqq^{2}\int_{0}^{\Lambda}d\ell \ell^{2}t_{0,Nt\to Nt}(k,q)t_{0,Nt\to Nt}(k,\ell)\\\nonumber
&\hspace{1cm}\times\frac{1}{\sqrt{\frac{3}{4}q^{2}-M_{N}E-i\epsilon}-\gamma_{t}}\frac{1}{\sqrt{\frac{3}{4}\ell^{2}-M_{N}E-i\epsilon}-\gamma_{t}}\frac{1}{q\ell}Q_{0}\left(\frac{-q^{2}-\ell^{2}-\lambda^{2}}{2q\ell}\right)\\\nonumber
&-\frac{3\rho_{s}\alpha M_{N}}{4\pi^{2}}\int_{0}^{\Lambda}\!\!dqq^{2}\int_{0}^{\Lambda}d\ell \ell^{2}t_{0,Nt\to Ns}(k,q)t_{0,Nt\to Ns}(k,\ell)\\\nonumber
&\hspace{1cm}\times\frac{1}{\sqrt{\frac{3}{4}q^{2}-M_{N}E-i\epsilon}-\gamma_{s}}\frac{1}{\sqrt{\frac{3}{4}\ell^{2}-M_{N}E-i\epsilon}-\gamma_{s}}\frac{1}{q\ell}Q_{0}\left(\frac{-q^{2}-\ell^{2}-\lambda^{2}}{2q\ell}\right).
\end{align}
While the partial resummation technique~\cite{Gabbiani:1999yv} can be used to calculate NLO $pd$ scattering, the result will also include a subset of higher order diagrams.  In that technique the LO+NLO scattering amplitude is calculated by using the integral equation for the LO scattering amplitude but (i) replacing all LO dibaryon propagators by LO+NLO dibaryon propagators; and (ii) modifying the Coulomb inhomogeneous term $\mathbf{B}_{0}^{(C)}(k,p,E)$ and kernel $\mathbf{K}_{0}^{(C)}(q,p,E)$ to include new contributions from photon exchanges between a dibaryon and a nucleon line.  These new contributions are given by
\begin{align}
\label{eq:inhomDoubletCNLO}
&\mathbf{B}^{(C)}_{1}(k,p)=\left(\begin{array}{c}
-\frac{\alpha \rho_{t} M_{N}y_{t}^{2}}{kp}Q_{0}\left(\frac{-k^{2}-p^{2}-\lambda^{2}}{2kp}\right)\\
0\\
0
\end{array}\right),
\end{align}
and
\begin{equation}
\label{eq:homDoubletCNLO}
\mathbf{K}^{(C)}_{1}(q,p,E)=
-\frac{\alpha M_{N}^{2}}{16\pi}\frac{1}{qp}Q_{0}\left(\frac{-q^{2}-p^{2}-\lambda^{2}}{2qp}\right)\left(\begin{array}{ccc}
y_{t}^{2}\rho_{t} & 0 & 0 \\
0 &  y_{s}^{2}\rho_{s} & 0\\
0 & 0 & 0 
\end{array}\right)\mathbf{D}^{(0)}(E-\frac{\vect{q}^{2}}{2M_{N}},\vect{q}).
\end{equation}
The true advantages of the partial resummation technique become apparent at NNLO, where it yields the straightforward computation of diagrams without having to calculate the full off-shell scattering amplitude.  However, a new technique has been developed that provides a strictly perturbative calculation of diagrams, also without the need to separately calculate full off-shell scattering amplitudes, and which is no more numerically expensive than the partial resummation technique~\cite{Vanasse:2013sda}.  Here we will consider both a strictly perturbative and partial resummation calculation of the NLO $pd$ scattering amplitudes.

\section{Expressions for Phase Shifts and Bound State Energies}

The physical elastic scattering amplitude $T_{0}(k)$ at LO is obtained by putting the scattering amplitude full on-shell $(k=p, E=\frac{3k^{2}}{4M_{N}}-\frac{\gamma_{t}^{2}}{M_{N}})$ and then multiplying by the LO deuteron wavefunction renormalization, yielding
\begin{align}
\label{eq:LOScattAmp}
T_{0}(k)=Z_{LO}t_{0,Nt\to Nt}\left(k,k,\frac{3k^{2}}{4M_{N}}-\frac{\gamma_{t}^{2}}{M_{N}}\right).
\end{align}
The NLO correction to the elastic scattering amplitude $T_{1}(k)$ is then obtained as
\begin{align}
\label{eq:NLOScattAmp}
T_{1}(k)=Z_{NLO}t_{0,Nt\to Nt}\left(k,k,\frac{3k^{2}}{4M_{N}}-\frac{\gamma_{t}^{2}}{M_{N}}\right)+Z_{LO}t_{1,Nt\to Nt}\left(k,k,\frac{3k^{2}}{4M_{N}}-\frac{\gamma_{t}^{2}}{M_{N}}\right),
\end{align}
where $Z_{NLO}$ is the NLO correction to the deuteron wavefunction renormalization.  Both orbital and spin angular momenta are separately conserved at NLO in \EFT, so the scattering matrix can be decomposed into a completely diagonal basis of orbital and spin angular momenta.  Since the scattering matrix must be unitary, it has the following form in terms of a phase shift for the doublet S-wave channel:
\begin{equation}
\label{eq:Smatrix}
S=e^{2i\delta}.
\end{equation}
The scattering matrix is related to the scattering amplitude $T(k)$ via
\begin{equation}
\label{eq:STmatrix}
S=1+i\frac{2M_{N}k}{3\pi}T(k).
\end{equation}
Expanding both Eqs.~(\ref{eq:Smatrix}) and~(\ref{eq:STmatrix}) perturbatively yields
\begin{equation}
\label{eq:PhaseShiftLO}
\delta_{0}(k)=\frac{1}{2i}\ln\left(1+i\frac{2M_{N}k}{3\pi}T_{0}(k)\right),
\end{equation}
for the LO phase shift, and
\begin{equation}
\delta_{1}(k)=\frac{1}{2i}\frac{i\frac{2M_{N}k}{3\pi}T_{1}(k)}{1+i\frac{2M_{N}k}{3\pi}T_{0}(k)},
\end{equation}
for the NLO correction to the phase shift. 

For $pd$ scattering we use the Coulomb-subtracted phase shift, 
\begin{equation}
\delta_{n,\mathrm{diff}}(k)=\delta_{n,\mathrm{full}}(k)-\delta_{n,C}(k),
\end{equation}
where $\delta_{0,\mathrm{full}}(k)$ is the LO phase shift calculated by including all of the strong ($S$), strong-Coulomb ($SC$), and Coulomb ($C$) pieces in the integral equations.  The $\delta_{0,C}(k)$ phase shift is calculated by only including the Coulomb ($C$) pieces in the LO integral equations.  In this case, all three channels decouple, leaving a single channel integral equation to solve at LO.  The NLO correction, $\delta_{1,\mathrm{full}}(k)$, to the phase shift is obtained with the LO amplitude that again contains all ($S$), ($SC$), and ($C$) pieces.  This LO amplitude is then used with Eqs.~(\ref{eq:NLOt1}), (\ref{eq:NLOt2}), and~(\ref{eq:NLOt3}) to calculate the NLO amplitude.  For the NLO correction $\delta_{1,C}(k)$ the LO amplitude is calculated only using the ($C$) pieces.  Then this LO amplitude with only Eq.~(\ref{eq:NLOt3}) yields the NLO ``Coulomb'" amplitude. 

 In the partial resummation technique $\delta_{1,\mathrm{full}}(k)$ includes all of the ($S$), ($SC$), and ($C$) pieces as well as the additional kernel Eq.~(\ref{eq:homDoubletCNLO}) and inhomogeneous term Eq.~(\ref{eq:inhomDoubletCNLO}) in the integral equation.  For $\delta_{1,C}(k)$ in the partial resummation technique, only the ($C$) terms as well as the additional kernel Eq.~(\ref{eq:homDoubletCNLO}) and inhomogeneous term Eq.~(\ref{eq:inhomDoubletCNLO}) are kept in the integral equation.  Thus, the integral equations decouple again, leaving only a single channel integral equation.

In addition to $pd$ scattering we investigate the bound state properties of ${}^{3}\skHe$.  In particular, we want to be able to predict its binding energy.  At LO this is done by dropping the inhomogeneous term in the integral equation, leading to the homogeneous equation
\begin{equation}
\mathbf{t}_{0}(k,p,E)=\mathbf{K}_{0}(q,p,E)\otimes\mathbf{t}_{0}(k,q,E).
\end{equation}
This equation is essentially an eigenvalue problem with eigenvector $\mathbf{t}_{0}(k,q,E)$ and eigenvalue one.  Thus, the LO bound state energy $B_{0}$ is the energy for which
\begin{equation}
\det(1-\mathbf{K}_{0}(q,p,B_{0}))=0.
\end{equation}
The NLO correction to the bound state energy is calculated perturbatively.  We extend the method used by Ji and Phillips~\cite{Ji:2012nj} to include complications from isospin.  At the bound state energy the scattering amplitude possesses a pole and can be written
\begin{equation}
\mathbf{t}_{0}(k,p,E)+\mathbf{t}_{1}(k,p,E)+\cdots=\frac{\mathbf{Z}_{0}(k,p)+\mathbf{Z}_{1}(k,p)}{E+B_{0}+B_{1}}+\mathbf{R}_{0}(k,p,E)+\mathbf{R}_{1}(k,p,E)+\cdots,
\end{equation}
where $\mathbf{Z}_{0}(k,p)$ ($\mathbf{Z}_{1}(k,p)$) is the LO (NLO) smooth residue vector function about the pole, $\mathbf{R}_{0}(k,p,E)$ ($\mathbf{R}_{1}(k,p,E)$) the LO (NLO) smooth remainder vector function, and $B_{0}$ and $B_{1}$ are the LO binding energy and NLO correction to the binding energy, respectively.  Expanding this expression perturbatively and collecting all LO terms gives
\begin{equation}
\label{eq:LORes}
\mathbf{Z}_{0}(k,p)=\lim_{\scriptscriptstyle E\to -B_{0}}(E+B_{0})\mathbf{t}_{0}(k,p,E).
\end{equation}
Doing the same at NLO gives 
\begin{equation}
\label{eq:NLOenergy}
B_{1}=-\!\!\lim_{ \scriptscriptstyle E\to -B_{0}}\frac{(E+B_{0})^{2}[\mathbf{t}_{1}]_{n}(k,p,E)}{[\mathbf{Z}_{0}]_{n}(k,p)},
\end{equation}
or
\begin{equation}
B_{1}=-\!\!\lim_{\scriptscriptstyle E\to -B_{0}}\frac{(E+B_{0})^{2}\mathbf{Z}_{0}^{T}(k,p)\mathbf{t}_{1}(k,p,E)}{\mathbf{Z}_{0}^{2}(k,p)},
\end{equation}
for the NLO correction to the bound state energy. The subscript $n$ refers to any component of the three vector.  For $B_{1}$ the choice of $k$ and $p$ should be completely arbitrary.  This can be shown rigorously by first noting that the components of the LO residue vector function $\mathbf{Z}_{0}(k,p)$ can be factorized as~\cite{PhysRev.141.902,Ji:2011qg}
\begin{align}
\label{eq:Resfac}
\mathbf{Z}_{0}(k,p)=\left(\begin{array}{c}
\Gamma_{Nt}(k)\Gamma_{Nt}(p)\\
\Gamma_{Nt}(k)\Gamma_{Ns}(p)\\
\Gamma_{Nt}(k)\Gamma_{Npp}(p)
\end{array}\right).
\end{align}
The functions $\Gamma_{Nt}(p)$, $\Gamma_{Ns}(p)$, and $\Gamma_{Npp}(p)$ are components of the solution to the LO homogeneous integral equation $\boldsymbol{\Gamma}_{0}(p)$, which is given by
\begin{equation}
\label{eq:LOGamma}
\boldsymbol{\Gamma}_{0}(p)=\mathbf{K}_{0}(q,p,B_{0})\otimes\boldsymbol{\Gamma}_{0}(q),
\end{equation}
with $\boldsymbol{\Gamma}_{0}(p)$ defined in terms of its components as
\begin{equation}
\boldsymbol{\Gamma}_{0}(p)=\left(\begin{array}{c}
\Gamma_{Nt}(p) \\
\Gamma_{Ns}(p) \\
\Gamma_{Npp}(p) 
\end{array}\right).
\end{equation}
Note that the normalization of the LO homogeneous equation is not given by Eq.~(\ref{eq:LOGamma}) but can be obtained from Eq.~(\ref{eq:LORes}) or by using the techniques outlined in Ref.~\cite{Konig:2011yq}.  Substituting Eqs.~(\ref{eq:NLOt1}), (\ref{eq:NLOt2}), and~(\ref{eq:NLOt3}) for $[\mathbf{t}_{1}]_{1}(k,p,E)$ in Eq.~(\ref{eq:NLOenergy}) and using Eq.~(\ref{eq:LORes}) together with Eq.~(\ref{eq:Resfac}), all LO amplitudes occurring in $[\mathbf{t}_{1}]_{1}(k,p,E)$ and $[\mathbf{Z}_{0}]_{1}(k,p)$ are changed to products of components of the homogeneous equation after taking the limit.  The resulting expression for $B_{1}$ no longer has any dependence on the momenta $k$ and $p$ ($i\epsilon$ has been dropped because $E<0$ and all resulting square roots are positive):
\begin{align}
\label{eq:NLOB1}
B_{1}=&\frac{\rho_{t}}{4\pi}\int_{0}^{\Lambda}\!\!\!dq q^{2}\left(\Gamma_{Nt}(q)\right)^{2}\frac{\sqrt{\frac{3}{4}q^{2}-M_{N}E}+\gamma_{t}}{\sqrt{\frac{3}{4}q^{2}-M_{N}E}-\gamma_{t}}\\\nonumber
&+\frac{3\rho_{s}}{4\pi}\int_{0}^{\Lambda}\!\!\!dq q^{2}\left(\Gamma_{Ns}(q)\right)^{2}\frac{\frac{3}{4}q^{2}-M_{N}E}{\left(\sqrt{\frac{3}{4}q^{2}-M_{N}E}-\gamma_{s}\right)^{2}}\\\nonumber
&+\frac{3r_{C}}{8\pi}\int_{0}^{\Lambda}\!\!\!dq q^{2}\left(\Gamma_{Npp}(q)\right)^{2}\frac{\frac{3}{4}q^{2}-M_{N}E}{\left(2\kappa H\left(\frac{\kappa}{\sqrt{\frac{3}{4}q^{2}-M_{N}E}}\right)+\frac{1}{a_{C}}\right)^{2}}\\\nonumber
&+\frac{(H_{0,1}(\Lambda)+H_{0,1}^{(\alpha)}(\Lambda))}{\pi^{2}\Lambda^{2}}\vast[\int_{0}^{\Lambda}\!\!\!dq q^{2}\Gamma_{Nt}(q)\frac{1}{\sqrt{\frac{3}{4}q^{2}-M_{N}E}-\gamma_{t}}\\\nonumber
&+\int_{0}^{\Lambda}\!\!\!dq q^{2}\Gamma_{Ns}(q)\frac{1}{\sqrt{\frac{3}{4}q^{2}-M_{N}E}-\gamma_{s}}+\int_{0}^{\Lambda}\!\!\!dq q^{2}\Gamma_{Npp}(q)\frac{1}{-\frac{1}{a_{C}}-2\kappa H\left(\frac{\kappa}{\sqrt{\frac{3}{4}q^{2}-M_{N}E}}\right)}\vast]^{2}\\\nonumber
&-\frac{\rho_{t}\alpha M_{N}}{4\pi^{2}}\int_{0}^{\Lambda}\!\!\!dqq^{2}\int_{0}^{\Lambda}\!\!\!d\ell\ell^{2}\Gamma_{Nt}(q)\Gamma_{Nt}(\ell)\frac{1}{\sqrt{\frac{3}{4}q^{2}-M_{N}E}-\gamma_{t}}\\\nonumber
&\hspace{2cm}\times\frac{1}{\sqrt{\frac{3}{4}\ell^{2}-M_{N}E}-\gamma_{t}}\frac{1}{q\ell}Q_{0}\left(\frac{-q^{2}-\ell^{2}-\lambda^{2}}{2q\ell}\right)\\\nonumber
&-\frac{3\rho_{s}\alpha M_{N}}{4\pi^{2}}\int_{0}^{\Lambda}\!\!\!dqq^{2}\int_{0}^{\Lambda}\!\!\!d\ell\ell^{2}\Gamma_{Ns}(q)\Gamma_{Ns}(\ell)\frac{1}{\sqrt{\frac{3}{4}q^{2}-M_{N}E}-\gamma_{s}}\\\nonumber
&\hspace{2cm}\times\frac{1}{\sqrt{\frac{3}{4}\ell^{2}-M_{N}E}-\gamma_{s}}\frac{1}{q\ell}Q_{0}\left(\frac{-q^{2}-\ell^{2}-\lambda^{2}}{2q\ell}\right).
\end{align}

\section{Leading-Order Asymptotics: No New Counterterm at LO}
\label{sec:LOAsymptotics}

Observables must be independent of the momentum cutoff used to regulate the theory.  In particular, as $\Lambda\to\infty$, $\mathcal{O}(\frac{1}{\Lambda^{2}})$, etc. pieces are suppressed and the prediction should stabilize.  For the case of LO $nd$ scattering it is well established that a three-body force is required to obtain cutoff-independent results~\cite{Bedaque:1999ve}.  However, it has not been explicitly shown for the case of LO $pd$ scattering that no additional three-body force term is needed to remove possible additional cutoff-dependence introduced by the inclusion of the Coulomb diagrams that are necessary to describe $pd$ interactions.  Calculations of doublet-channel $pd$ scattering have been carried out in \EFT~\cite{Konig:2011yq}, but at LO these calculations did not go to sufficiently high cutoffs to definitively settle the question.  Here we show that there is no new LO three-body force required for $pd$ scattering.
In order to investigate the asymptotic behavior of the LO scattering amplitude we redefine the scattering amplitudes as
\begin{equation}
t_{+}(k,p)=t_{0,Nt\to Nt}(k,p)+t_{0,Nt\to Ns}(k,p)+t_{0,Nt\to Npp}(k,p),
\end{equation}
\begin{equation}
t_{-}(k,p)=t_{0,Nt\to Nt}(k,p)-t_{0,Nt\to Ns}(k,p)-t_{0,Nt\to Npp}(k,p),
\end{equation}
and
\begin{equation}
t_{\emptyset}(k,p)=t_{0,Nt\to Ns}(k,p)-\frac{1}{2}t_{0,Nt\to Npp}(k,p).
\end{equation}
In addition we define the dibaryon propagators
\begin{equation}
D_{+}(E,\vect{q})=\left(y_{t}^{2}D_{t}^{(0)}(E,\vect{q})+\frac{1}{3}y_{s}^{2}D_{s}^{(0)}(E,\vect{q})+\frac{2}{3}y_{s}^{2}D_{pp}^{(0)}(E,\vect{q})\right),
\end{equation}
\begin{equation}
D_{-}(E,\vect{q})=\left(y_{t}^{2}D_{t}^{(0)}(E,\vect{q})-\frac{1}{3}y_{s}^{2}D_{s}^{(0)}(E,\vect{q})-\frac{2}{3}y_{s}^{2}D_{pp}^{(0)}(E,\vect{q})\right),
\end{equation}
and
\begin{equation}
D_{\emptyset}(E,\vect{q})=y_{s}^{2}\left(D_{s}^{(0)}(E,\vect{q})-D_{pp}^{(0)}(E,\vect{q})\right).
\end{equation}
The LO scattering amplitude is still given by Eq.~(\ref{eq:LOScatt}).  However, the definition of the vector $\mathbf{t}_{0}(k,p)$ is now replaced by
\begin{equation}
\label{eq:newt}
\mathbf{t}_{0}(k,p)=\left(\begin{array}{c}
t_{+}(k,p)\\
t_{-}(k,p)\\
t_{\emptyset}(k,p)
\end{array}\right).
\end{equation}
Likewise $\mathbf{K}^{(S)}_{0}(q,p,E)$ becomes 
\begin{align}
\label{eq:Kmod}
\mathbf{K}^{(S)}_{0}(q,p,E)=&\frac{M_{N}}{8\pi}\frac{1}{qp}Q_{0}\left(\frac{q^{2}+p^{2}-M_{N}E-i\epsilon}{qp}\right)\left(\begin{array}{ccc}
-2D_{+} &  -2D_{-} & -\frac{8}{3}D_{\emptyset} \\
D_{-} & D_{+} & -\frac{4}{3}D_{\emptyset} \\
\frac{1}{3}D_{\emptyset} & -\frac{1}{3}D_{\emptyset} & D_{+}-D_{-}+\frac{2}{3}D_{\emptyset}
\end{array}\right)\\\nonumber
&+\frac{M_{N}}{8\pi}\frac{2H_{0,0}(\Lambda)}{\Lambda^{2}}\left(\begin{array}{ccc}
-D_{+} & -D_{-} & -\frac{4}{3}D_{\emptyset}\\
0 & 0 & 0 \\
0 & 0 & 0 
\end{array}\right),
\end{align}
where $D_{+}$, $D_{-}$, and $D_{\emptyset}$ are dibaryon propagators with energy and momentum arguments $(E-\frac{\vect{q}^{2}}{2M_{N}},\vect{q})$.  To study the asymptotic limit ($q\sim p \gg \Lambda_{\not{\pi}}$) of the amplitudes, we keep only terms up to $\mathcal{O}(1/\Lambda^{2})$, yielding
\begin{align}
\label{eq:KSCmod}
&\mathbf{K}_{0}^{(SC)}(q,p,E)=\frac{M_{N}}{16\pi}\left(\begin{array}{ccc}
\frac{2}{3}(C(q,p,E)+V_{2}(q,p,E)+V_{1}(q,p,E)) & 0 & 0\\
\frac{1}{3}(C(q,p,E)-2V_{2}(q,p,E)+V_{1}(q,p,E)) & 0 & 0\\
\frac{1}{6}(C(q,p,E)-2V_{2}(q,p,E)+V_{1}(q,p,E)) & 0 & 0\\
\end{array}\right)D_{+}+\cdots,
\end{align}
(see again Eqs.~(\ref{eq:Vertex})--(\ref{eq:Cross})) and, 
\begin{align}
\label{eq:KCmod}
\mathbf{K}_{0}^{(C)}(q,p,E)=
\frac{M_{N}}{16\pi}\left(\begin{array}{ccc}\vspace{-.2cm}
1 & 0 & 0\\\vspace{-.2cm}
\nicefrac{1}{2} & 0 & 0\\
\nicefrac{1}{4} & 0 & 0\\
\end{array}\right)\frac{1}{3}B(q,p,E)D_{+}+\cdots,
\end{align}
(see again Eq.~(\ref{eq:Bubble})), using the newly defined amplitudes.  There is no need to redefine $\mathbf{B}_{0}(q,p,E)$ because it is suppressed in the asymptotic limit.  The terms that have been omitted in the definitions of $\mathbf{K}^{SC}(q,p,E)$ and $\mathbf{K}^{(C)}(q,p,E)$ will become important for higher orders in the \EFT expansion.  The dibaryon propagators expanded in the asymptotic limit yield
\begin{subequations}
\begin{equation}
\label{eq:D+def}
D_{+}(E,\vect{q})\sim-\frac{4\pi}{M_{N}}\left(2\sqrt{\frac{4}{3}}\frac{1}{q}+\frac{4}{3}\left(\gamma_{t}+\frac{1}{3}\gamma_{s}+\frac{2}{3}\gamma_{C}\right)\frac{1}{q^{2}}+\frac{16}{9}\frac{\kappa\ln(q)}{q^{2}}\right)+\cdots,
\end{equation}
\begin{equation}
D_{-}(E,\vect{q})\sim-\frac{4\pi}{M_{N}}\left( \frac{4}{3}\left(\gamma_{t}-\frac{1}{3}\gamma_{s}-\frac{2}{3}\gamma_{C}\right)\frac{1}{q^{2}}-\frac{16}{9}\frac{\kappa\ln(q)}{q^{2}}\right)+\cdots,
\end{equation}
and
\begin{equation}
\label{eq:D0def}
D_{\emptyset}(E,\vect{q})\sim-\frac{4\pi}{M_{N}}\left( \frac{4}{3}\left(\gamma_{s}-\gamma_{C}\right)\frac{1}{q^{2}}-\frac{8}{3}\frac{\kappa\ln(q)}{q^{2}}\right)+\cdots,
\end{equation}
\end{subequations}
where $\gamma_{C}$ is defined as
\begin{equation}
\gamma_{C}=\frac{1}{a_{C}}-2C_{E}\kappa-2\kappa\ln\left(\sqrt{\frac{4}{3}}\kappa\right),
\end{equation}
with $C_{E}\simeq0.5772$ the Euler--Mascheroni constant.  The scattering amplitude in the asymptotic limit is obtained by using Eqs.~(\ref{eq:newt})-(\ref{eq:KCmod}) and  Eqs.~(\ref{eq:D+def})-(\ref{eq:D0def}) in Eq.~(\ref{eq:LOScatt}).  Then, using appropriate ans{\"a}tze (see appendix for details), the asymptotic behavior of the scattering amplitudes can be obtained.  The resulting asymptotic forms are
\begin{multline}
\label{eq:t+}
t_{+}(q)=C\left\{\frac{\sin\left(s_{0}\ln\left(\frac{q}{\Lambda^{*}}\right)\right)}{q}+\frac{1}{\sqrt{3}}\left(\gamma_{t}+\frac{1}{3}\gamma_{s}+\frac{2}{3}\gamma_{C}\right)|B_{-1}|\sin\frac{\left(s_{0}\ln\left(\frac{q}{\Lambda^{*}}\right)+\mathrm{Arg}(B_{-1})\right)}{q^2}\right.\\
+\frac{4\kappa}{3\sqrt{3}}|C_{-1}|\frac{\sin\left(s_{0}\ln\left(\frac{q}{\Lambda^{*}}\right)+\mathrm{Arg}(C_{-1})\right)}{q^2}+\frac{4\kappa}{3\sqrt{3}}|D_{-1}|\frac{\sin\left(s_{0}\ln\left(\frac{q}{\Lambda^{*}}\right)+\mathrm{Arg}(D_{-1})\right)}{q^2}\\
\left.+\frac{4\kappa}{3\sqrt{3}}|B_{-1}|\ln(q)\frac{\sin\left(s_{0}\ln\left(\frac{q}{\Lambda^{*}}\right)+\mathrm{Arg}(B_{-1})\right)}{q^2}\right.\\
-\left.\frac{16\kappa}{\sqrt{3}\pi}|E_{-1}|\frac{\sin\left(s_{0}\ln\left(\frac{q}{\Lambda^{*}}\right)+\mathrm{Arg}(E_{-1})\right)}{q^{2}}+\cdots\right\},
\end{multline}
\begin{multline}
\label{eq:t-}
t_{-}(q)=C\left\{-\frac{1}{2\sqrt{3}}\left(\gamma_{t}-\frac{1}{3}\gamma_{s}-\frac{2}{3}\gamma_{C}\right)|\tilde{B}_{-1}|\frac{\sin\left(s_{0}\ln\left(\frac{q}{\Lambda^{*}}\right)+\mathrm{Arg}(\tilde{B}_{-1})\right)}{q^{2}}\right.\\
+\frac{2\kappa}{3\sqrt{3}}|\tilde{C}_{-1}|\frac{\sin\left(s_{0}\ln\left(\frac{q}{\Lambda^{*}}\right)+\mathrm{Arg}(\tilde{C}_{-1})\right)}{q^{2}}-\frac{\kappa}{3\sqrt{3}}|\tilde{D}_{-1}|\frac{\sin\left(s_{0}\ln\left(\frac{q}{\Lambda^{*}}\right)+\mathrm{Arg}(\tilde{D}_{-1})\right)}{q^{2}}\\
\left.+\frac{2\kappa}{3\sqrt{3}}|\tilde{B}_{-1}|\ln(q)\frac{\sin\left(s_{0}\ln\left(\frac{q}{\Lambda^{*}}\right)+\mathrm{Arg}(\tilde{B}_{-1})\right)}{q^{2}}\right.\\
-\left.\frac{16\kappa}{\sqrt{3}\pi}|\tilde{E}_{-1}|\frac{\sin\left(s_{0}\ln\left(\frac{q}{\Lambda^{*}}\right)+\mathrm{Arg}(\tilde{E}_{-1})\right)}{q^{2}}+\cdots\right\},
\end{multline}
and,
\begin{multline}
\label{eq:t0}
t_{\emptyset}(q)=C\left\{-\frac{1}{6\sqrt{3}}\left(\gamma_{s}-\gamma_{C}\right)|\tilde{B}_{-1}|\frac{\sin\left(s_{0}\ln\left(\frac{q}{\Lambda^{*}}\right)+\mathrm{Arg}(\tilde{B}_{-1})\right)}{q^{2}}\right.\\
+\frac{\kappa}{3\sqrt{3}}|\tilde{C}_{-1}|\frac{\sin\left(s_{0}\ln\left(\frac{q}{\Lambda^{*}}\right)+\mathrm{Arg}(\tilde{C}_{-1})\right)}{q^{2}}-\frac{\kappa}{6\sqrt{3}}|\tilde{D}_{-1}|\frac{\sin\left(s_{0}\ln\left(\frac{q}{\Lambda^{*}}\right)+\mathrm{Arg}(\tilde{D}_{-1})\right)}{q^{2}}\\
\left.+\frac{\kappa}{3\sqrt{3}}|\tilde{B}_{-1}|\ln(q)\frac{\sin\left(s_{0}\ln\left(\frac{q}{\Lambda^{*}}\right)+\mathrm{Arg}(\tilde{B}_{-1})\right)}{q^{2}}\right.\\
-\left.\frac{8\kappa}{\sqrt{3}\pi}|\tilde{E}_{-1}|\frac{\sin\left(s_{0}\ln\left(\frac{q}{\Lambda^{*}}\right)+\mathrm{Arg}(\tilde{E}_{-1})\right)}{q^{2}}+\cdots\right\}.
\end{multline}

Note that the leading asymptotic form for $t_{+}(q)$ is exactly the same as in $nd$ scattering~\cite{Bedaque:2002yg}.  However, the subleading $t_{-}(q)$, $t_{\emptyset}(q)$, and the subleading part of $t_{+}(q)$ are modified in $pd$ scattering.  In addition to acquiring a $\ln(q)$ piece, these amplitudes receive electromagnetic corrections in $\kappa=\frac{\alpha M_{N}}{2}$ and isospin breaking effects from $\gamma_{C}\neq\gamma_{s}$.  The asymptotic form of these amplitudes in $nd$ scattering is obtained by setting $\kappa=0$ and $\gamma_{s}=\gamma_{C}$.  This asymptotic form can also be obtained by replacing $D_{pp}(E,q)$ by $D_{s}(E,q)$.  In this limit, $D_{\emptyset}(E,q)=0$, and in Eq.~(\ref{eq:Kmod}) $t_{\emptyset}(q)$ decouples from $t_{+}(q)$ and $t_{-}(q)$.  This  leaves two coupled integral equations.  In the Wigner $SU(4)$ limit~\cite{PhysRev.51.106,Mehen:1999qs} ($\gamma_{t}=\gamma_{s}$), the $t_{+}(q)$ and $t_{-}(q)$ equations decouple.  The resulting equation for $t_{+}(q)$ is equivalent to a 
three-boson problem and has a well-known solution that requires a three-body 
force to obtain cutoff-independent results~\cite{Bedaque:1998km}.  The equation for $t_{-}(q)$ in this limit is equivalent to $nd$ scattering in the quartet S-wave channel and does not require a three-body force for cutoff-independence.

\begin{figure}[hbt]
\includegraphics[width=100mm]{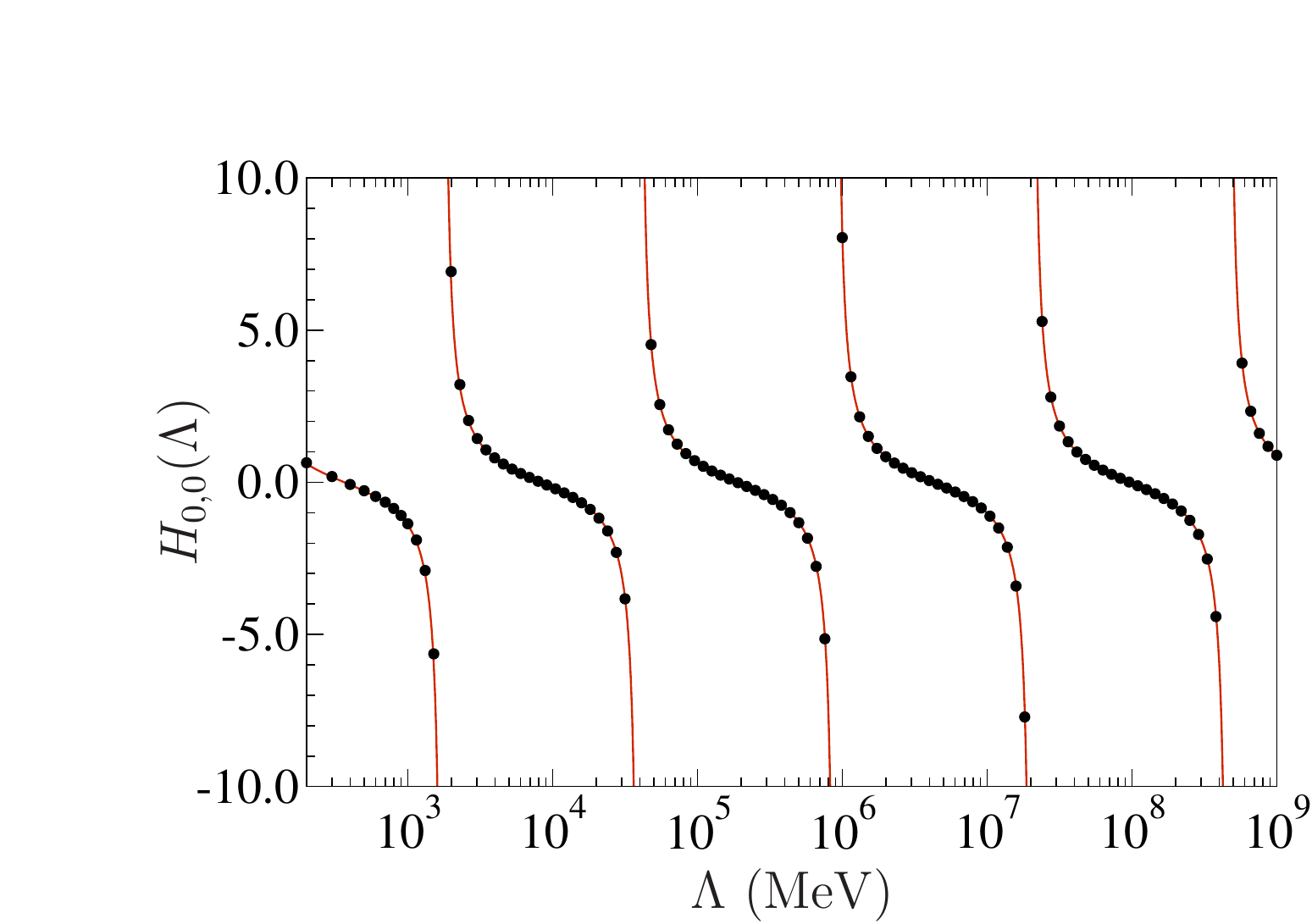}
\caption{\label{fig:H0}(Color online) Comparison of numerical and analytical calculations of LO three-body force for $nd$ scattering, with $c\simeq0.877$ and $\Lambda^{*}\simeq 1.55$~MeV.  The three-body force is numerically fit to give the correct doublet S-wave $nd$ scattering length, $a_{n-d}=0.65$~fm.}
\end{figure}

An analytical approximation for the LO three-body force is obtained by plugging the asymptotic form of the scattering amplitudes into Eq.~(\ref{eq:LOScatt}), keeping the three-body force in the homogeneous term, and then demanding that the results are cutoff-independent to order $1/\Lambda$.  It is only necessary to keep the leading $t_{+}(q)$ amplitude when considering cutoff independence to order $1/\Lambda$, and because the leading behavior of $t_{+}(q)$ is the same in both $nd$ and $pd$ scattering, the LO three-body force is the same in both cases. Its approximate analytic form is~\cite{Bedaque:1998km}
\begin{equation}
H_{0,0}(\Lambda)=c\frac{\sin\left(s_{0}\ln\left(\frac{\Lambda}{\Lambda^{*}}\right)+\arctan{s_{0}}\right)}{\sin\left(s_{0}\ln\left(\frac{\Lambda}{\Lambda^{*}}\right)-\arctan{s_{0}}\right)},
\end{equation}
where $c$ is a regulator-dependent quantity.  For our choice of cutoff regularization we find $c=0.877\pm0.003$ fits the numerical results to the analytical form.  Within the error of our fit for $c$, we find good agreement with previous results~\cite{Braaten:2011sz}.  Fitting to the numerical data yields $\Lambda^{*}\simeq 1.55$ MeV.  This value is not exactly the same as the $\Lambda^{*}\simeq 1.63$ MeV found in the equations for the asymptotic amplitudes (see Eqs.~(\ref{eq:t+})-(\ref{eq:t0}) and Sec.~\ref{sec:NLO-3B-Phases}), most likely due to finite-$\Lambda$ effects.  The results of matching the numerical and analytical results for $H_{0,0}(\Lambda)$ are shown in Fig.~\ref{fig:H0}, where $H_{0,0}(\Lambda)$ is fit to give the correct doublet S-wave $nd$ scattering length, $a_{n-d}=0.65$~fm.  Agreement between the numerical and analytical results is clear.
\begin{figure}[hbt!]
\includegraphics[width=100mm]{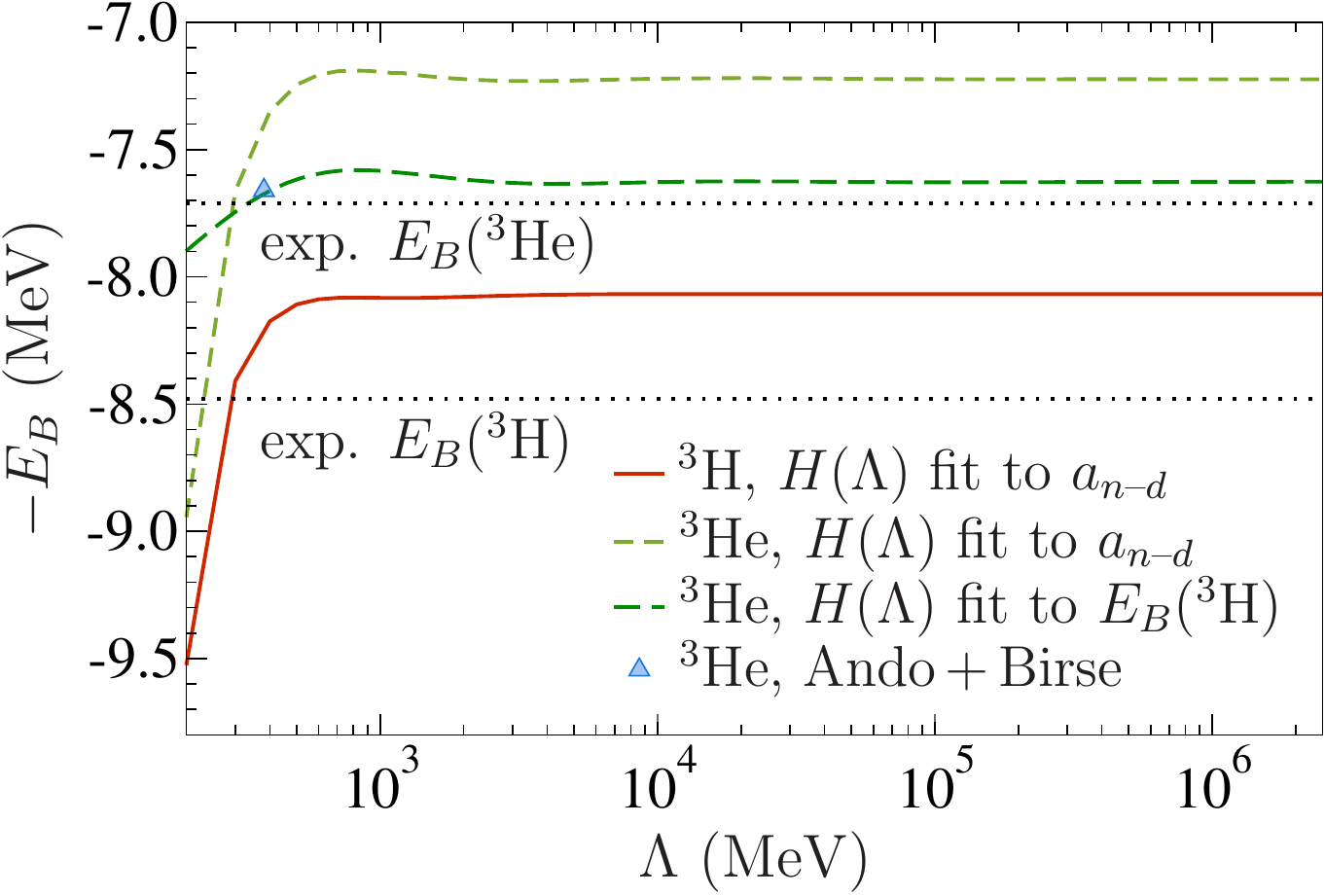}
\caption{\label{fig:Binding}(Color online) Cutoff dependence of LO predictions for ${}^{3}\skHe$ and ${}^{3}\skH$ binding energies.  The solid line is the ${}^{3}\skH$ binding energy prediction when the LO three-body force is fit to the $nd$ doublet S-wave scattering length.  The short dashed line is the ${}^{3}\skHe$ binding energy predicted when the the LO three-body force is fit to the $nd$ doublet S-wave scattering length. The long dashed line is the ${}^{3}\skHe$ binding energy prediction when the LO three-body force is fit to the ${}^{3}\skH$ binding energy.  Finally, the triangle point is a LO \EFT calculation by Ando and Birse in which Coulomb effects are treated nonperturbatively~\cite{Ando:2010wq}.}
\end{figure}

Using the same three-body force $H_{0,0}(\Lambda)$ to calculate the LO binding energies of ${}^{3}\skH$ and ${}^{3}\skHe$ yields the cutoff dependence shown in Fig.~\ref{fig:Binding}.  The binding energies clearly converge as a function of the cutoff used in the integral equation.  In Fig.~\ref{fig:Binding}, the solid line is the LO ${}^{3}\skH$ binding energy prediction when the three-body force is numerically fit to the $nd$ doublet S-wave scattering length, and the short-dashed line the LO ${}^{3}\skHe$ binding energy prediction using the same three-body force.  Fitting the three-body force to the ${}^{3}\skH$ binding energy of $B_{{}^{3}\skH}= 8.481798 \pm 0.000002$~MeV yields the LO ${}^{3}\skHe$ binding energy prediction given by the long-dashed line.  The triangle point is a \EFT calculation by Ando and Birse of the LO ${}^{3}\skHe$ binding energy at a cutoff of $\Lambda=380.689$ MeV, where Coulomb effects are treated fully nonperturbatively~\cite{Ando:2010wq}.  The long-dashed line essentially 
passes through the triangle point, which confirms that within the bound state regime of ${}^{3}\skHe$ it is a good approximation to treat Coulomb effects perturbatively.

Using the three-body force $H_{0,0}(\Lambda)$ fit to the doublet S-wave $nd$ scattering length we analyze the cutoff dependence of the LO $pd$ S-wave phase shift from 200 to $10^{7}$~MeV and find good convergence as the cutoff is increased.
These results are shown in Fig.~\ref{fig:shiftreal} in Sec.~\ref{sec:NLO-3B-Phases}.  The cutoff independence in both the $pd$ phase shifts and ${}^{3}\skHe$ binding energies confirms numerically that $H_{0,0}(\Lambda)$ is the only three-body force needed at LO for both $nd$ and $pd$ scattering.
%
%
\section{NLO Behavior without New $pd$ Counterterm}

\begin{figure}[hbt]
\includegraphics[width=100mm]{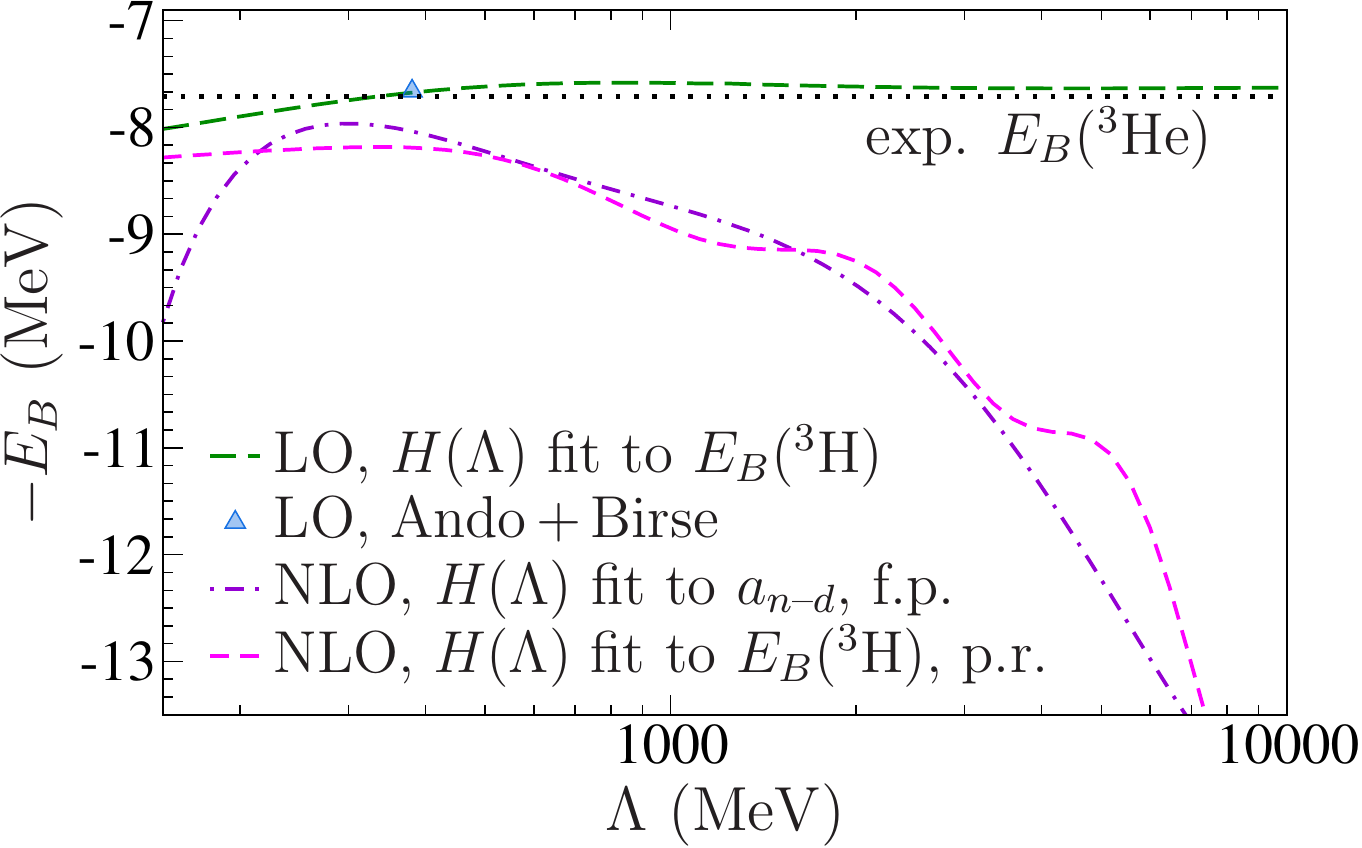}
\caption{\label{fig:B1n0H1a}(Color online) The long-dashed line is the LO ${}^{3}\skHe$ binding energy prediction when the LO three-body force is fit to the ${}^{3}\skH$ binding energy.  The short dashed line is the NLO ${}^{3}\skHe$ binding energy prediction in the partial resummation technique when the three-body force is to fit the ${}^{3}\skH$ binding energy~\cite{Koenig:2013,Konig:2014ufa}.  The dashed-dotted line is the NLO ${}^{3}\skHe$ binding energy in a strictly perturbative approach for the case where the LO and NLO $nd$ three-body force is fit to the doublet S-wave $nd$ scattering length and any possible new $pd$ three-body force is ignored.  Finally, the triangle point is again the LO \EFT calculation by Ando and Birse~\cite{Ando:2010wq}.}
\end{figure}

To address if a new $pd$ three-body force is needed at NLO in addition to the NLO $nd$ three-body force, we can calculate the cutoff dependence of various physical quantities.  The NLO ${}^{3}\skHe$ binding energy results are shown in Fig.~\ref{fig:B1n0H1a}.
The result using the partial resummation technique is given by the short-dashed line, where a LO+NLO three-body force is fit to the ${}^{3}\skH$ binding energy. (A more detailed analysis of the
 partial-resummation calculation can be
 found in Ref.~\cite{Konig:2014ufa}; for a preliminary discussion, among other things,
 see also Ref.~\cite{Koenig:2013}.) The NLO ${}^{3}\skHe$ binding energy is clearly diverging for higher cutoffs.  The dashed-dotted line in Fig.~\ref{fig:B1n0H1a} is the ${}^{3}\skHe$ binding energy prediction in a strictly perturbative calculation, where the LO and NLO three-body forces are both separately fit to reproduce the $nd$ doublet S-wave scattering length $a_{n-d}=0.65$~fm.  Again, for larger cutoffs the binding energy prediction is clearly diverging; a new NLO $pd$ three-body force is needed to make these results independent of the cutoff.  In the next section we derive an expression for this three-body force and demonstrate that indeed it gives cutoff-independent phase 
shifts.

\section{NLO Three-body Forces and Predicted Phase Shifts}
\label{sec:NLO-3B-Phases}

To obtain an approximate analytical form for the NLO three-body forces $H_{0,1}(\Lambda)$ and $H_{0,1}^{(\alpha)}(\Lambda)$, we begin with the NLO correction to the ${}^{3}\skHe$ binding energy, Eq.~(\ref{eq:NLOB1}).  Redefining the solution to the homogeneous equation as
\begin{equation}
\Gamma_{+}(q)=\Gamma_{Nt}(q)+\Gamma_{Ns}(q)+\Gamma_{Npp}(q),
\end{equation}
\begin{equation}
\Gamma_{-}(q)=\Gamma_{Nt}(q)-\Gamma_{Ns}(q)-\Gamma_{Npp}(q),
\end{equation}
and
\begin{equation}
\Gamma_{\emptyset}(q)=\Gamma_{Ns}(q)-\frac{1}{2}\Gamma_{Npp}(q),
\end{equation}
is entirely analogous to the redefinition of the LO scattering amplitudes used to analyze the LO asymptotic behavior. In fact, the asymptotic solutions for the scattering amplitudes $t_{+}(k,q)$,  $t_{-}(k,q)$, and $t_{\emptyset}(k,q)$ are also the asymptotic solutions for $\Gamma_{+}(q)$,  $\Gamma_{-}(q)$, and $\Gamma_{\emptyset}(q)$, respectively.  Using this redefinition of the homogeneous equation with Eq.~(\ref{eq:NLOB1}), plugging in the asymptotic solutions Eqs.~(\ref{eq:t+})-(\ref{eq:t0}), using dimensional analysis, and keeping only those terms that diverge in the UV limit ($\Lambda\to\infty$) yields
\begin{align}
\label{eq:B1div}
&[B_{1}]_{\text{UV-div}}=\frac{1}{4\pi}\left(\frac{1}{4}\rho_{t}+\frac{1}{12}\rho_{s}+\frac{1}{6}r_{C}\right)\int^{\Lambda}\!\!\!dq q^{2}(\Gamma_{+}^{(-1)}(q))^{2}\\\nonumber
&+\frac{1}{2\pi}\left(\frac{1}{4}\rho_{t}+\frac{1}{12}\rho_{s}+\frac{1}{6}r_{C}\right)\int^{\Lambda}\!\!\!dq
q^{2}\Gamma_{+}^{(-1)}(q)\Gamma_{+}^{(-2)}(q)\\\nonumber
&+\frac{1}{4\pi}\left(\frac{1}{2}\rho_{t}-\frac{1}{6}\rho_{s}-\frac{1}{3}r_{C}\right)\int^{\Lambda}\!\!\!dq q^{2}\Gamma_{+}^{(-1)}(q)\Gamma_{-}^{(-2)}(q)+\frac{1}{4\pi}\left(\frac{2}{3}\rho_{s}-\frac{2}{3}r_{C}\right)\int^{\Lambda}dq q^{2}\Gamma_{+}^{(-1)}(q)\Gamma_{\emptyset}^{(-2)}(q)\\\nonumber
&+\frac{1}{\sqrt{3}\pi}\left(\frac{1}{4}\rho_{t}\gamma_{t}+\frac{1}{12}\rho_{s}\gamma_{s}+\frac{1}{6}r_{C}\gamma_{C}\right)\int^{\Lambda}\!\!\!dqq(\Gamma_{+}^{(-1)}(q))^{2}+\frac{\kappa r_{C}}{3\sqrt{3}\pi}\int^{\Lambda}\!\!\!dqq\ln(q)(\Gamma_{+}^{(-1)}(q))^{2}\\\nonumber
&+\frac{4(H_{0,1}(\Lambda)+H_{0,1}^{(\alpha)}(\Lambda))}{3\pi^{2}\Lambda^{2}}\left\{\left(\int^{\Lambda}\!\!\!dqq\Gamma_{+}^{(-1)}(q)\right)^{2}+2\int^{\Lambda}\!\!\!dqq\Gamma_{+}^{(-1)}(q)\int^{\Lambda}\!\!\!d\ell\ell\Gamma_{+}^{(-2)}(\ell)\right.\\\nonumber
&+\sqrt{\frac{4}{3}}\left(\gamma_{t}+\frac{1}{3}\gamma_{s}+\frac{2}{3}\gamma_{C}\right)\int^{\Lambda}\!\!\!dqq\Gamma_{+}^{(-1)}(q)\int^{\Lambda}\!\!\!d\ell\Gamma_{+}^{(-1)}(\ell)\\\nonumber
&\left.+\sqrt{\frac{4}{3}}\frac{4\kappa}{3}\int^{\Lambda}\!\!\!dqq\Gamma_{+}^{(-1)}(q)\int^{\Lambda}d\ell\ln(\ell)\Gamma_{+}^{(-1)}(\ell)\right\}\\\nonumber
&-\frac{\alpha M_{N}}{3\pi^{2}}\left(\frac{1}{4}\rho_{t}+\frac{1}{12}\rho_{s}\right)\int^{\Lambda}\!\!dq\int^{\Lambda}\!\!d\ell \Gamma_{+}^{(-1)}(q)\Gamma_{+}^{(-1)}(\ell)Q_{0}\left(\frac{-q^{2}-\ell^{2}-\lambda^{2}}{2q\ell}\right),
\end{align}
where the superscript ``$(n)$'' on the homogeneous solutions refers to the $\mathcal{O}(\Lambda^{n})$ piece of the asymptotic solution.  To cancel the UV divergences, $H_{0,1}(\Lambda)$+$H_{0,1}^{(\alpha)}(\Lambda)$ must be chosen such that $[B_{1}]_{\text{UV-div}}$ is zero.  $H_{0,1}(\Lambda)$, the NLO three-body force for $nd$ scattering, is found by setting $\kappa\to0$, $r_{C}\to \rho_{s}$, and $\gamma_{C}\to\gamma_{s}$ and then solving the resulting equation, including all divergent and $\mathcal{O}(\Lambda^{0})$ pieces. (Since numerically $\ln(\Lambda)$ and $\ln^{2}(\Lambda)$ terms are $\mathcal{O}(\Lambda^{0})$ except for extremely large cutoffs~\cite{Ji:2011qg}, which are not considered here, we treat these terms as $\mathcal{O}(\Lambda^{0})$.)\comment{(Numerically $\ln(\Lambda)$ and $\ln^{2}(\Lambda)$ terms are $\mathcal{O}(\Lambda^{0})$ except for extremeley large cutoffs~\cite{Ji:2011qg}, which we do not consider, therefore we treat these terms as $\mathcal{O}(\Lambda^{0})$)}  The resulting 
expression for $H_{0,1}(\Lambda)$ has a linear divergence and $\mathcal{O}(\Lambda^{0})$ pieces that are given by 
\begin{align}
\label{eq:H1nd}
&H_{0,1}(\Lambda)=\Lambda h_{10}(\Lambda)-\frac{3\pi(1+s_{0}^{2})}{64}\left\{\frac{1}{\sqrt{3}}(\rho_{t}+\rho_{s})(\gamma_{t}+\gamma_{s})|B_{-1}|\mathcal{G}_{1}(B_{-1})\right.\\\nonumber
&\hspace{2cm}-\frac{1}{2\sqrt{3}}(\rho_{t}-\rho_{s})(\gamma_{t}-\gamma_{s})|\tilde{B}_{-1}|\mathcal{G}_{1}(\tilde{B}_{-1})\\\nonumber
&\hspace{2cm}\left.+\frac{2}{\sqrt{3}}(\rho_{t}\gamma_{t}+\rho_{s}\gamma_{s})\mathcal{G}_{1}(0)+f\right\}/\sin^{2}\left(s_{0}\ln\left(\frac{\Lambda}{\Lambda^{*}}\right)-\arctan(s_{0})\right),
\end{align}
where,
\begin{equation}
\mathcal{G}_{1}(x)=\cos(\mathrm{Arg}(x))\ln(\Lambda)-\frac{1}{2s_{0}}\sin\left(2s_{0}\ln\left(\frac{\Lambda}{\Lambda^{*}}\right)+\mathrm{Arg}(x)\right),
\end{equation}
and the function $h_{10}(\Lambda)$ multiplying the linear divergence is
\begin{equation}
h_{10}(\Lambda)=-\frac{3\pi(1+s_{0}^{2})}{128}(\rho_{t}+\rho_{s})\frac{\left(1-\frac{1}{\sqrt{1+4s_{0}^{2}}}\sin\left(2s_{0}\ln\left(\frac{\Lambda}{\Lambda^{*}}\right)+\arctan\left(\frac{1}{2s_{0}}\right)\right)\right)}{\sin^{2}\left(s_{0}\ln\left(\frac{\Lambda}{\Lambda^{*}}\right)-\arctan(s_{0})\right)}.
\end{equation}
A previous calculation of the $nd$ three-body force $H_{0,1}(\Lambda)$ appeared in Ref.~\cite{Hammer:2001gh}.  However, this calculation dropped the contribution from the linear divergence.  In addition, the authors did not include additional subleading terms and isospin-breaking terms.  Despite this, their numerical results for the phase shifts are still correct as they numerically fit their NLO three-body force to the $nd$ scattering length.  In the exact isospin limit, $\rho=\rho_{t}=\rho_{s}$ and $\gamma=\gamma_{t}=\gamma_{s}$, Eq.~(\ref{eq:H1nd}) reduces to that of Ref.~\cite{Ji:2011qg}.  However, our solution does not contain the piece with a triple pole as in their result.  This is because we do not explicitly split $H_{0,1}(\Lambda)$ into two pieces; unlike their calculation, our scattering length is always fixed.  The value $f$ in Eq.~(\ref{eq:H1nd}) contains the details of the infrared (IR) regularization of the integrals.  The value of $f$ depends on the regularization scheme and renormalization 
condition and its value is obtained by fitting to the numerical data of the three-body force, $H_{0,1}(\Lambda)$.  At sufficiently large 
cutoffs the value of $f$ is irrelevant; the linear divergence will dominate over this $\mathcal{O}(\Lambda^{0})$ term.
For convenience we split up the three-body force term $H_{0,1}^{(\alpha)}(\Lambda)$ as
\begin{equation}\label{eq:h1a_split}
H_{0,1}^{(\alpha)}(\Lambda)=h_{I}^{(\alpha)}(\Lambda)+h_{\kappa}^{(\alpha)}(\Lambda),
\end{equation}
where $h_{I}^{(\alpha)}(\Lambda)$ are contributions from isospin breaking and $h_{\kappa}^{(\alpha)}(\Lambda)$ are terms with an explicit $\kappa$ from electromagnetic effects.  (Actually, $\gamma_{C}$ contains $\kappa$ pieces, so a part of it should be relegated to $h_{\kappa}^{(\alpha)}(\Lambda)$.  However, we will include all the contributions of $\gamma_{C}$ in $h_{I}^{(\alpha)}(\Lambda)$ for convenience.)  Plugging Eqs.~(\ref{eq:H1nd})~and~(\ref{eq:h1a_split}) into Eq.~(\ref{eq:B1div}) and keeping all divergent and $\mathcal{O}(\Lambda^{0})$ terms provides the three-body force terms
\begin{align}
\label{eq:HalphaI}
&h_{I}^{(\alpha)}(\Lambda)=-\frac{3\pi(1+s_{0}^{2})}{16}\times\\\nonumber
&\left\{\frac{1}{12}(r_{C}-\rho_{s})\Lambda\left[1-\frac{1}{\sqrt{1+4s_{0}^{2}}}\sin\left(2s_{0}\ln\left(\frac{\Lambda}{\Lambda^{*}}\right)+\arctan\left(\frac{1}{2s_{0}}\right)\right)\right]\right.\\\nonumber
&+\frac{1}{3\sqrt{3}}\left(\frac{1}{2}(\rho_{t}+\rho_{s})(\gamma_{C}-\gamma_{s})+\frac{1}{2}(r_{C}-\rho_{s})(\gamma_{t}+\gamma_{s})+\frac{1}{3}(r_{C}-\rho_{s})(\gamma_{C}-\gamma_{s})\right)|B_{-1}|\mathcal{G}_{1}(B_{-1})\\\nonumber
&-\frac{1}{12\sqrt{3}}\left(\frac{4}{3}(r_{C}-\rho_{s})(\gamma_{C}-\gamma_{s})-(\rho_{t}-\rho_{s})(\gamma_{C}-\gamma_{s})-(r_{C}-\rho_{s})(\gamma_{t}-\gamma_{s})\right)|\tilde{B}_{-1}|\mathcal{G}_{1}(\tilde{B}_{-1})\\\nonumber
&+\frac{1}{3\sqrt{3}}(r_{C}\gamma_{C}-\rho_{s}\gamma_{t})\mathcal{G}_{1}(0)-\frac{64}{9\sqrt{3}\pi s_{0}\sqrt{1+s_{0}^{2}}}h_{10}(\Lambda)\sin\left(s_{0}\ln\left(\frac{\Lambda}{\Lambda^{*}}\right)-\arctan(s_{0})\right)(\gamma_{C}-\gamma_{s})\\\nonumber
&\times\left[|B_{_1}|\mathcal{G}_{3}(B_{-1})+\mathcal{G}_{3}(0)\right]+f\Bigg{\}}/\sin^{2}\left(s_{0}\ln\left(\frac{\Lambda}{\Lambda^{*}}\right)+\arctan(s_{0})\right),
\end{align}
and
\begin{align}
\label{eq:Halphakappa}
&h_{\kappa}^{(\alpha)}=-\frac{\sqrt{3}\kappa\pi(1+s_{0}^{2})}{48}\left\{\left(\rho_{t}+\frac{1}{3}\rho_{s}+\frac{2}{3}r_{C}\right)\left[\frac{}{}|C_{-1}|\mathcal{G}_{1}(C_{-1})+|D_{-1}|\mathcal{G}_{1}(D_{-1})\right.\right.\\\nonumber
&\left.-\frac{12}{\pi}|E_{-1}|\mathcal{G}_{1}(E_{-1})+\frac{1}{2}|B_{-1}|\mathcal{G}_{2}(B_{-1})\right]+r_{C}\mathcal{G}_{2}(0)\\\nonumber
&+\frac{1}{2}\left(\rho_{t}+\frac{1}{3}\rho_{s}-\frac{4}{3}r_{c}\right)\left[|\tilde{C}_{-1}|\mathcal{G}_{1}(\tilde{C}_{-1})-\frac{1}{2}|\tilde{D}_{-1}|\mathcal{G}_{1}(\tilde{D}_{-1})-\frac{24}{\pi}|\tilde{E}_{-1}|\mathcal{G}_{1}(\tilde{E}_{-1})+\frac{1}{2}|\tilde{B}_{-1}|\mathcal{G}_{2}(\tilde{B}_{-1})\right]\\\nonumber
&-\frac{128}{3\pi s_{0}\sqrt{1+s_{0}^{2}}}h_{10}(\Lambda)\sin\left(s_{0}\ln\left(\frac{\Lambda}{\Lambda^{*}}\right)-\tan^{-1}(s_{0})\right)\left[\frac{}{{}_{}}|C_{-1}|\mathcal{G}_{3}(C_{-1})+|D_{-1}|\mathcal{G}_{3}(D_{-1})\right.\\\nonumber
&\left.\left.-\frac{1}{s_{0}}|B_{-1}|\mathcal{G}_{4}(B_{-1})-12|E_{-1}|\mathcal{G}_{3}(E_{-1})-\frac{1}{s_{0}}\mathcal{G}_{4}(0)\right]+\Psi(\Lambda)\right\}/\\\nonumber
&\sin^{2}\left(s_{0}\ln\left(\frac{\Lambda}{\Lambda^{*}}\right)+\arctan{s_{0}}\right).
\end{align}
The functions $\mathcal{G}_{2}(x)$, $\mathcal{G}_{3}(x)$, and $\mathcal{G}_{4}(x)$ are defined by 
\begin{align}
&\mathcal{G}_{2}(x)=\cos(\mathrm{Arg}(x))\ln^{2}(\Lambda)-\frac{1}{2s_{0}^{2}}\cos\left(2s_{0}\ln\left(\frac{\Lambda}{\Lambda^{*}}\right)+\mathrm{Arg}(x)\right)\\\nonumber
&\hspace{5cm}-\frac{1}{s_{0}}\ln(\Lambda)\sin\left(2s_{0}\ln\left(\frac{\Lambda}{\Lambda^{*}}\right)+\mathrm{Arg}(x)\right),
\end{align}
\begin{equation}
\mathcal{G}_{3}(x)=\cos\left(s_{0}\ln\left(\frac{\Lambda}{\Lambda^{*}}\right)+\mathrm{Arg}(x)\right),
\end{equation}
and
\begin{equation}
\mathcal{G}_{4}(x)=\sin\left(s_{0}\ln\left(\frac{\Lambda}{\Lambda^{*}}\right)+\mathrm{Arg}(x)\right)-s_{0}\ln(\Lambda)\cos\left(s_{0}\ln\left(\frac{\Lambda}{\Lambda^{*}}\right)+\mathrm{Arg}(B_{-1})\right).
\end{equation}
The function $\Psi(\Lambda)$ contains contributions where a photon is exchanged between a dibaryon and a nucleon, and is given by the double integral appearing in the last line of Eq.~(\ref{eq:B1div}).  To obtain the analytical form of the asymptotic behavior we fit to the following function
\begin{equation}
\Psi(\Lambda)=a\ln(\Lambda)+b\sin\left(2s_{0}\left(\frac{\Lambda}{\Lambda^{*}}\right)+c\right)+d.
\end{equation}
This form gives good agreement with the numerical data if $a=0.0234$, $b=-0.0153$, $c=0.579$, and $d=-0.0885$.  The value $f$ in Eq.~(\ref{eq:HalphaI}) again refers to the details of the IR regularization and depends on the regularization scheme and renormalization conditions.  The value of $f$ is determined by fitting to the numerical data for $H_{0,1}^{(\alpha)}(\Lambda)$. For the three-body force, $H_{0,1}^{(\alpha)}(\Lambda)$, we consider the two scenarios $r_{C}\neq\rho_{s}$ and $r_{C}=\rho_{s}$.  For $r_{C}\neq\rho_{s}$, $H_{0,1}^{(\alpha)}(\Lambda)$ has a linear divergence such that for sufficiently large cutoffs it will always dominate, making the value of $f$ unimportant there.  For $r_{C}=\rho_{s}$, the linear divergence disappears and the worst divergence is $\ln(\Lambda)^{2}$, which numerically is $\mathcal{O}(\Lambda^{0})$ except for extremely large cutoffs.  Since $f$ is also $\mathcal{O}(\Lambda^{0})$ it has a more sizable impact when $r_{C}=\rho_{s}$.

\begin{figure}[hbt]
\includegraphics[width=100mm]{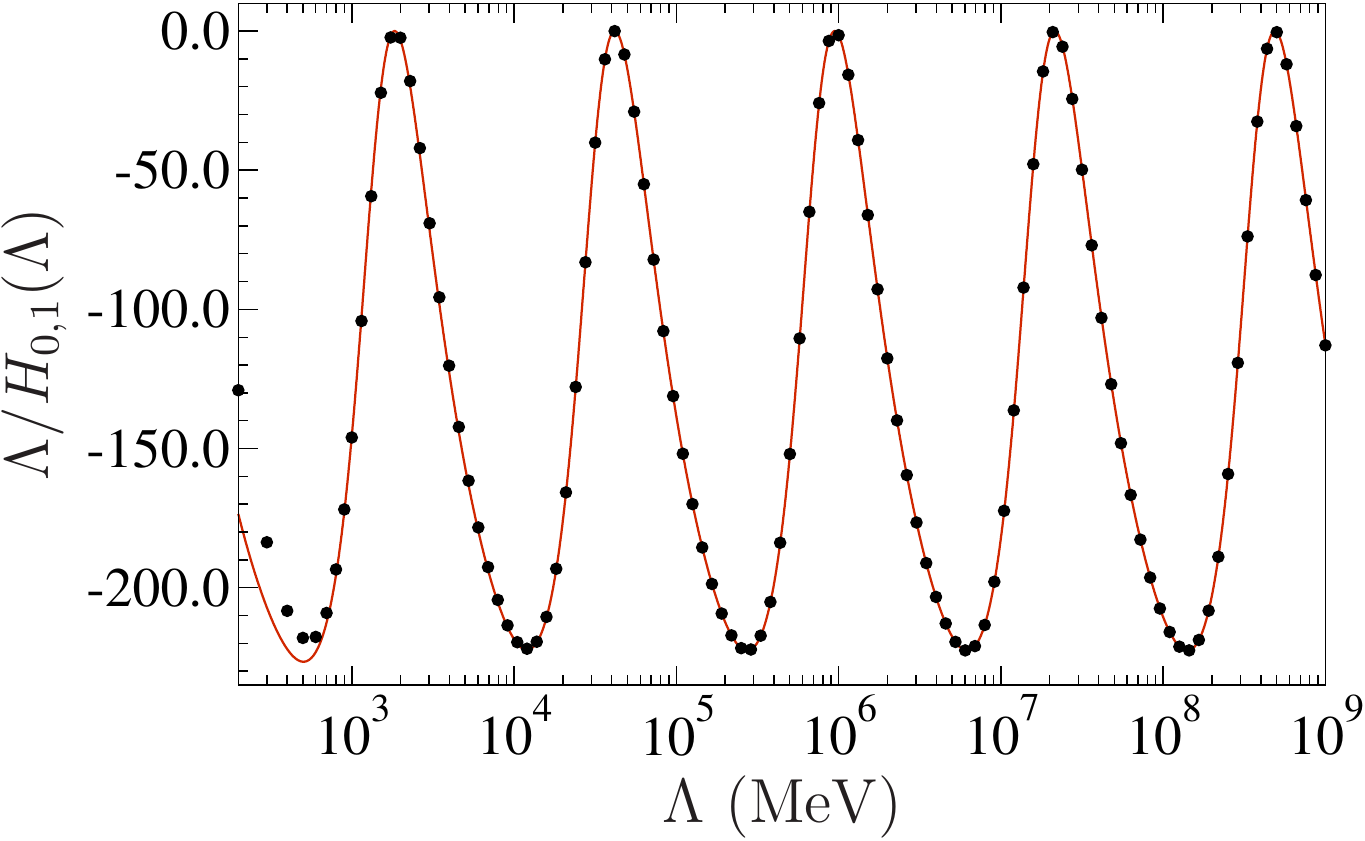}
\caption{\label{fig:H1}(Color online) NLO $nd$ three-body force comparison of numerical and analytic calculations.  $H_{0,1}(\Lambda)$ is fixed to reproduce the doublet S-wave $nd$ scattering length.  The value of $f$ for $H_{0,1}(\Lambda)$ is $f\simeq-0.1252$.}
\end{figure}

The analytical and numerical results for $\Lambda/H_{0,1}(\Lambda)$ are shown in Fig.~\ref{fig:H1}.  In $\Lambda/H_{0,1}(\Lambda)$, the dominant linear divergence is divided out and this form quickly asymptotes to a sinusoidal function.  Also, all poles of $H_{0,1}(\Lambda)$ are converted to zeroes.  The numerical results are obtained by fixing $H_{0,1}(\Lambda)$ to obtain the correct doublet S-wave $nd$ scattering length $a_{nd}=0.65$~fm.  To numerically determine $H_{0,1}(\Lambda)$ we calculate the LO $nd$ scattering amplitudes for a cutoff of $\Lambda=10^{12}$~MeV.  Then we use this LO amplitude to calculate the NLO integrals for smaller cutoffs up to $\bar{\Lambda}=10^{9}$~MeV.  This ensures that finite-$\Lambda$ effects are suppressed up to a factor of $\bar{\Lambda}/\Lambda=10^{-3}$.  The value $f$ for $H_{0,1}(\Lambda)$ is found to be $f\simeq -0.1252$.  For cutoffs below roughly 1000~MeV there seem to be notable discrepancies between the analytical and numerical predictions.  This is no surprise for 
$nd$ 
scattering since the asymptotic solution does not match the numerical solution for the LO scattering amplitude below cutoffs of about 1000~MeV.  The value of $\Lambda^{*}$ for all of the three-body forces is $\Lambda^{*}\simeq 1.63$ MeV, which is exactly the same $\Lambda^{*}$ appearing in the asymptotic form of the LO scattering amplitudes, Eqs.~\eqref{eq:t+}-\eqref{eq:t0}.

\begin{figure}[hbt]
\includegraphics[width=100mm]{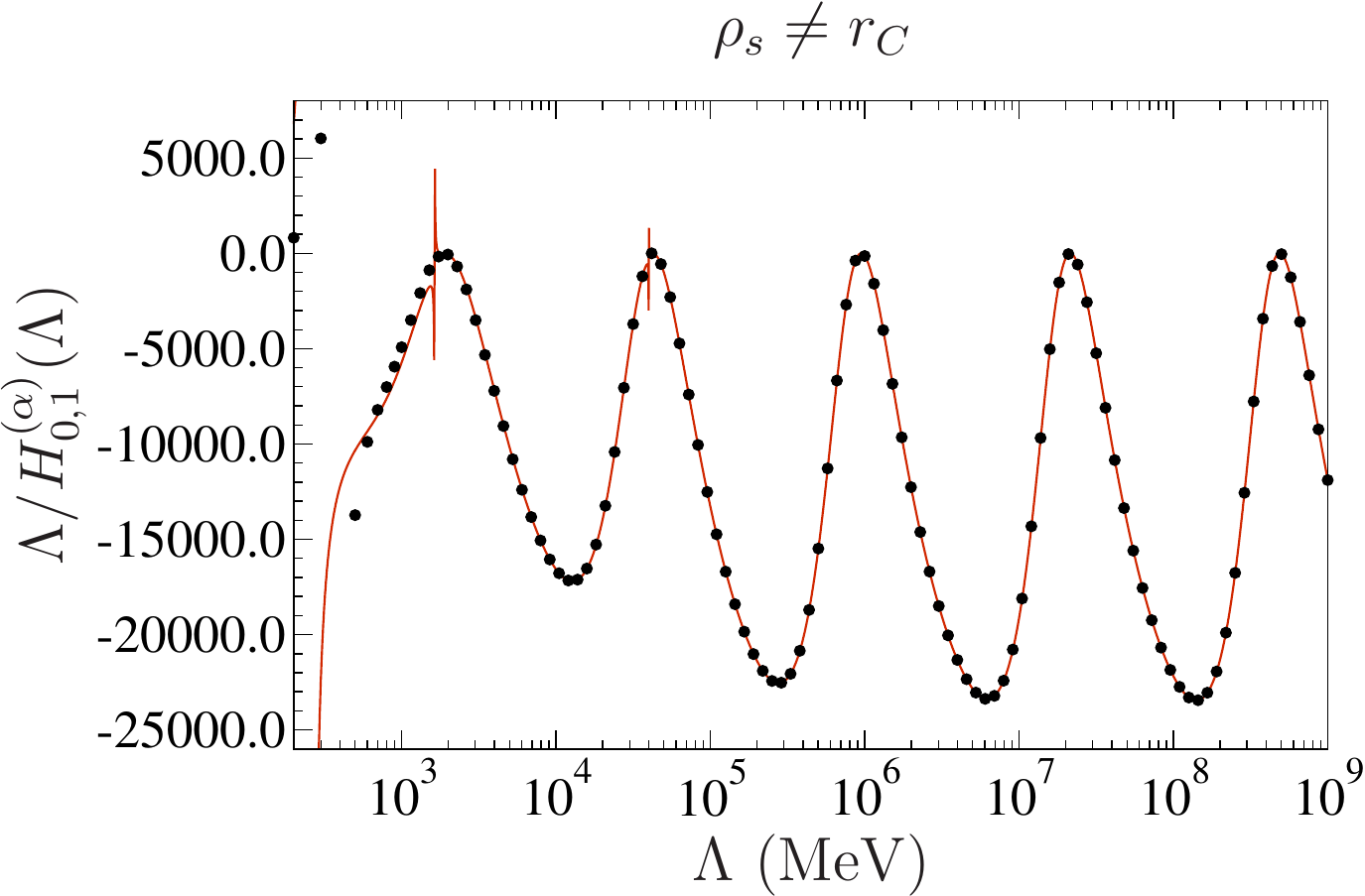}
\caption{\label{fig:H1ar0tnotrc}(Color online) NLO $pd$ three-body force comparison of numerical and analytic calculations for $r_{c}\neq\rho_{s}$. $H_{0,1}^{(\alpha)}(\Lambda)$ is fixed to give the correct ${}^{3}\skHe$ binding energy at NLO.  The value of $f$ for $H_{0,1}^{(\alpha)}(\Lambda)$ is $f\simeq0.1570$.}
\end{figure}

The numerical and analytical results for $\Lambda/H_{0,1}^{(\alpha)}(\Lambda)$ with $r_{C}\neq \rho_{s}$ are shown in Fig.~\ref{fig:H1ar0tnotrc}.  The three-body force $H_{0,1}^{(\alpha)}(\Lambda)$ is fixed to reproduce the physical ${}^{3}\skHe$ binding energy $-7.718043\pm0.000002$~MeV at NLO.  Using $B_{1}$ from either Eq.~(\ref{eq:NLOenergy}) or Eq.~(\ref{eq:NLOB1}) gives equivalent results.  We calculate either the homogeneous solution or the scattering amplitude up to a cutoff $\Lambda=10^{12}$~MeV, depending on which equation is used to determine $B_{1}$.  Then we calculate the NLO integrals for $B_{1}$ for either equation up to a cutoff of $\bar{\Lambda}=10^{9}$~MeV.  This is again to suppress finite-$\Lambda$ effects.  The value of $f$ for $H_{0,1}^{(\alpha)}(\Lambda)$ is found to be $f\simeq0.1570$.  Below roughly 5000~MeV we find notable differences between the numerical and analytical predictions.  This again is not surprising as for $pd$ scattering asymptotic solutions to the 
amplitude do not match the numerical solution below cutoffs of 5000~MeV. 

\begin{figure}[hbt]
\includegraphics[width=100mm]{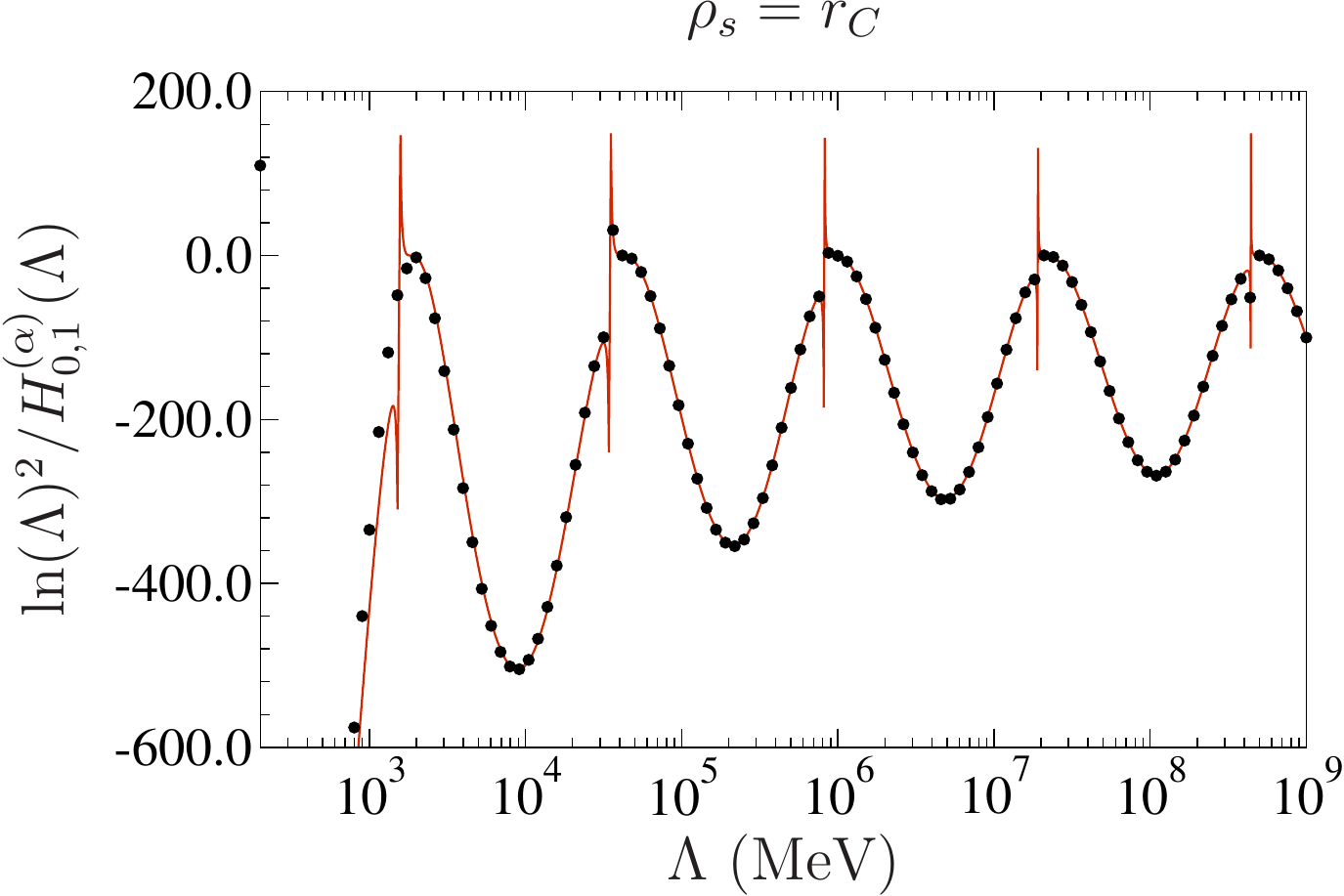}
\caption{\label{fig:H1ar0tisrc}(Color online) NLO $pd$ three-body force comparison of numerical and analytic calculations for $r_{c}=\rho_{s}$. $H_{0,1}^{(\alpha)}(\Lambda)$ is fixed to give the correct ${}^{3}\skHe$ binding energy at NLO.  The value of $f$ for $H_{0,1}^{(\alpha)}(\Lambda)$ is $f\simeq0.1503$.}
\end{figure}

The numerical and analytical results for $\ln(\Lambda)^{2}/H_{0,1}^{(\alpha)}(\Lambda)$ with $r_{C}=\rho_{s}$ are given in Fig.~\ref{fig:H1ar0tisrc}.  Again this choice divides away the somewhat dominant $\ln(\Lambda)^{2}$ dependence and converts all poles of $H_{0,1}^{(\alpha)}(\Lambda)$ to zeroes.  The three-body force $H_{0,1}^{(\alpha)}(\Lambda)$ is calculated in exactly the same manner as for the case $r_{C}\neq\rho_{s}$, except in our NLO integrals we set $r_{C}=\rho_{s}$.  The value of $f$ for $H_{0,1}^{(\alpha)}(\Lambda)$ in this case is $f\simeq 0.1503$.  Again for $\Lambda\lesssim 5000$~MeV there are notable differences between the numerical and analytical results and the reason for such disagreement is the same as in the $r_{C}\neq \rho_{s}$ case.  Also, the triple pole from the term with $h_{10}(\Lambda)$ in Eqs.~(\ref{eq:HalphaI}) and~(\ref{eq:Halphakappa}), leading to the observed spikes, is more dominant since the linear divergence is absent.
\begin{figure}[hbt]
\includegraphics[width=100mm]{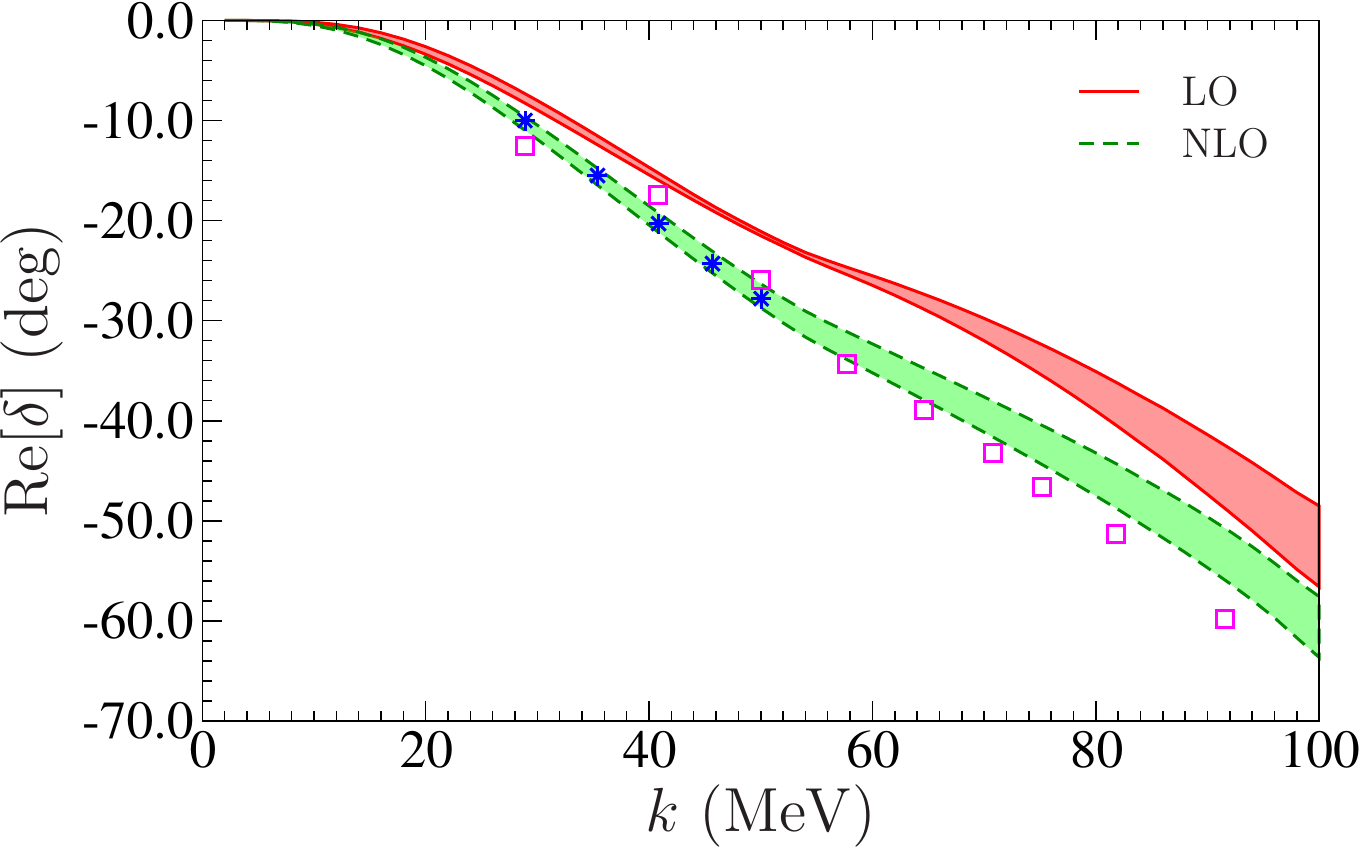}
\caption{\label{fig:shiftreal}(Color online) LO and NLO $pd$ scattering phase shifts.  The band comes from varying the cutoff from 200 to $10^7$~MeV.  The star points come from an AV-18 potential calculation~\cite{Kievsky:1996ca} and the open squares from a $pd$ phase shift analysis~\cite{Arvieux:1974}.}
\end{figure}

With the three-body forces fixed we now calculate the LO and NLO phase shifts in $pd$ scattering, as shown in Fig.~\ref{fig:shiftreal}.  The bands in the plot are generated by varying the cutoff from 200 to $10^7$~MeV.  With the new $H_{0,1}^{(\alpha)}(\Lambda)$ three-body force added there is clear convergence in the NLO phase shifts.  If $H_{0,1}^{(\alpha)}(\Lambda)$ is removed convergence is no longer observed.  At NLO these results are expected to have a roughly 17\% uncertainty ($\gamma_{t}^{2}\rho_{t}^{2}\sim 17\%$).  The star points come from calculations with the AV-18 potential and wavefunctions determined using the hyperspherical harmonic method~\cite{Kievsky:1996ca}; they agree with our results within the 17\% uncertainty.  The open squares come from a $pd$ phase shift analysis~\cite{Arvieux:1974} and also agree with our results within this uncertainty.

\section{Conclusion}

By analyzing the asymptotic form of the $pd$ scattering amplitude we have shown explicitly that at LO no new three-body force is needed for $pd$ scattering beyond those for $nd$ scattering.  This has been confirmed numerically by showing that the LO ${}^{3}\skHe$ binding energy and $pd$ scattering phase shifts are cutoff-independent using only the LO three-body force from $nd$ scattering.  In the three-body sector we included only electromagnetic terms that arise from iterating single Coulomb photon exchanges.  Based on the power counting of diagrams, treating Coulomb effects fully nonperturbatively as in Ref.~\cite{Ando:2010wq} should not change this result.

Using the asymptotic form of the $nd$ scattering amplitude we derived an analytical expression for the NLO $nd$ three-body force.  In the exact isospin limit our results agree with previous findings~\cite{Ji:2011qg}.  However, our results disagree with those of other authors~\cite{Hammer:2001gh} because they dropped linearly divergent terms and some subleading pieces.  Numerically calculating the NLO $nd$ three-body force by fitting to the doublet S-wave $nd$ scattering length, $a_{n-d}$, we find good agreement with our analytical form.   Using only the NLO $nd$ three-body force to calculate the NLO ${}^{3}\skHe$ binding energy and $pd$ scattering phase shift, both strictly perturbatively and using the partial resummation technique, does not produce cutoff-stable results, clearly indicating the need for a new $pd$ three-body force.

From the asymptotic form of the $pd$ scattering amplitude we have calculated an analytical form for this new $pd$ three-body force, $H_{0,1}^{(\alpha)}(\Lambda)$.  Calculating $H_{0,1}^{(\alpha)}(\Lambda)$ numerically by fixing the NLO correction to give the correct ${}^{3}\skHe$ binding energy gives good agreement between the analytical and numerical forms, both for the case $r_{c}\neq \rho_{s}$ and for $r_{c}=\rho_{s}$.  Finally, using the new $pd$ three-body force we obtain cutoff-independent NLO phase shifts for $pd$ scattering.

At NNLO in doublet S-wave $nd$ scattering there is a NNLO correction to the $H_{0}(\Lambda)$ three-body force and an additional new energy-dependent three-body force~\cite{Bedaque:2002yg,Ji:2012nj}.  In the case of $pd$ scattering these three-body forces will receive Coulomb and isospin-breaking corrections.  Thus $pd$ scattering at NNLO will very likely require two new three-body forces beyond those for $nd$ scattering that need to be renormalized to $pd$ and ${}^{3}\skHe$ data.  Possible renormalization conditions include fixing to the ${}^{3}\skHe$ binding energy and the $pd$ doublet S-wave scattering length, $a_{p-d}$.  Since this quantity is difficult to determine, it might be preferable to instead use other bound-state properties of ${}^{3}\skHe$, such as the charge radius.  We defer addressing these questions to future work, but note here that an NNLO \EFT calculation of $pd$ scattering will be an important first step towards understanding polarization asymmetries and in particular the $A_{y}$ problem~\cite{Huber:1998hu}.

\begin{acknowledgments}
 We would like to thank Harald
 W. Grie{\ss}hammer, Hans-Werner Hammer, U. van Kolck,
 and Daniel R. Phillips for useful discussions, and
 Thomas Mehen for comments on the manuscript.  We would also like to thank Matthias R. Schindler for pointing out an error. This work is supported in part by the Department of Energy under Grant No. DE-FG02-05ER41368, and by the NSF under Grant Nos.~PHY--1002478 and PHY--1306250.  Moreover, S.K. acknowledges support from the ``Studien\-stiftung des deutschen 
Volkes'' and the Bonn-Cologne Graduate School of Physics and Astronomy.  Also J.K. acknowledges support
from the Georgetown Hichwa Undergraduate Research Fellowship.
\end{acknowledgments}

\appendix

\section{Coulomb Diagrams}

The bubble diagram of Fig.~\ref{fig:CoulombLODiagrams}a (excluding spin and isospin dependence) is given by
\begin{align}
&-\frac{ie^{2}y_{t}^{2}M_{N}^{2}}{4\pi}\frac{1}{(\vect{k}-\vect{p})^{2}+\lambda^{2}}\frac{1}{|\vect{k}-\vect{p}|}\times\\\nonumber
&\times\tan^{-1}\left(\frac{|\vect{k}-\vect{p}|}{2\sqrt{\frac{3}{4}\vect{k}^{2}-M_{N}E-i\epsilon}+2\sqrt{\frac{3}{4}\vect{p}^{2}-M_{N}E-i\epsilon}}\right).
\end{align}
To perform the S-wave angular projection we make the substitution $z=|\vect{k}-\vect{p}|$, yielding
\begin{align}
&\frac{ie^{2}y_{t}^{2}M_{N}^{2}}{4\pi kp}\int_{|k-p|}^{k+p}dz\frac{1}{z^{2}+\lambda^{2}}\tan^{-1}\left(\frac{z}{2\sqrt{\frac{3}{4}\vect{k}^{2}-M_{N}E-i\epsilon}+2\sqrt{\frac{3}{4}\vect{p}^{2}-M_{N}E-i\epsilon}}\right)
\end{align}
This integral can be solved in terms of logarithms and dilogarithms, yielding Eq.~(\ref{eq:Bubble}) up to a constant factor from the spin and isospin dependence.

The vertex diagram of Fig.~\ref{fig:CoulombLODiagrams}b is given by the expression
\begin{align}
&\frac{ie^{2}y_{s}y_{j}M_{N}^{2}}{16\pi|\vect{k}+2\vect{p}|}\tan^{-1}\left(\frac{|\vect{k}+2\vect{p}|}{2\sqrt{\frac{3}{4}\vect{k}^{2}-M_{N}E-i\epsilon}+2\lambda}\right)\frac{1}{\vect{k}^{2}+\vect{p}^{2}+\vect{k}\cdot\vect{p}-M_{N}E-i\epsilon},
\end{align}
again without spin and isospin factors.  The substitution $z=|\vect{k}+2\vect{p}|$ gives the S-wave projection of the vertex diagram as
\begin{align}
&\frac{ie^{2}y_{s}y_{j}M_{N}^{2}}{8\pi kp}\int_{k+2p}^{|k-2p|}\frac{1}{z^{2}+3\vect{k}^{2}-4M_{N}E-i\epsilon}\tan^{-1}\left(\frac{z}{2\sqrt{\frac{3}{4}\vect{k}^{2}-M_{N}E-i\epsilon}+2\lambda}\right).\\\nonumber
\end{align}
This integral is similar to that of the bubble diagram and again can be solved in terms of logarithms and dilogarithms, yielding Eq.~(\ref{eq:Vertex}), again up to a constant spin-isospin factor.

The ``cross" diagram of Fig.~\ref{fig:CoulombLODiagrams}c can be written using Feynman parameters as
\begin{align}
&\frac{iy_{i}y_{j}e^{2}M_{N}^{2}}{16\pi}\int_{0}^{1}\!\!\!dx\int_{0}^{1}\!\!\!dy y\left[-y^{2}(\vect{p}+2\vect{k}+x(\vect{p}-\vect{k}))^{2}\right.\\\nonumber
&\left.\hspace{5cm}+4y(\vect{p}^{2}+\vect{k}^{2}+\vect{p}\cdot\vect{k}-M_{N}E-i\epsilon-\lambda^{2})+4\lambda^{2}\right]^{-\nicefrac{3}{2}}
\end{align}
This expression can in principle be solved exactly~\cite{Bedaque:1999vb}.  
However, the resulting form is too lengthy for practical numerical computation, 
so instead we expand in powers of $\lambda\ll\gamma_{t}$, 
yielding~\cite{Hoferichter-BoxDiag:2010}
\begin{align}
&\frac{iy_{i}y_{j}e^{2}M_{N}^{2}}{16\pi}\left\{\frac{1}{\vect{p}^{2}+\vect{k}^{2}+\vect{k}\cdot\vect{p}-M_{N}E-i\epsilon}\frac{1}{|\vect{k}-\vect{p}|}\times\right.\\\nonumber
&\hspace{3cm}\tan^{-1}\left(\frac{|\vect{k}-\vect{p}|}{2\sqrt{\frac{3}{4}\vect{p}^{2}-M_{N}E-i\epsilon}+2\sqrt{\frac{3}{4}\vect{k}^{2}-M_{N}E-i\epsilon}}\right)\\\nonumber
&\hspace{4cm}\left.-\frac{1}{2}\frac{\lambda}{(\vect{p}^{2}+\vect{k}^{2}+\vect{k}\cdot\vect{p}-M_{N}E-i\epsilon)^{2}}+\mathcal{O}(\lambda^{2})+\cdots\right\}.
\end{align}
The $\mathcal{O}(\lambda^{0})$ term is like the bubble diagram and its angular projection in the S-wave can be carried out similarly.  The $\mathcal{O}(\lambda^{1})$ term has a trivial S-wave angular projection.  Combining both angular projections we find Eq.~(\ref{eq:Cross}) up to a constant from spin and isospin projections.

\section{Asymptotics}

Collecting all terms to $\mathcal{O}(\Lambda^{-2})$ from Eqs.~\ref{eq:Kmod}--\ref{eq:KCmod}, $t_{+}(p)$ asymptotically is given by the integral equation
\begin{align}
t_{+}(p)=&\frac{4}{\sqrt{3}\pi}\frac{1}{p}\int_{0}^{\infty}\!\!dq\ln\left(\frac{q^{2}+pq+p^{2}}{q^{2}-pq+p^{2}}\right)t_{+}(q)\\\nonumber
&+\frac{4}{3\pi}\left(\gamma_{t}+\frac{1}{3}\gamma_{s}+\frac{2}{3}\gamma_{C}\right)\frac{1}{p}\int_{0}^{\infty}dq\ln\left(\frac{q^{2}+pq+p^{2}}{q^{2}-pq+p^{2}}\right)\frac{1}{q}t_{+}(q)\\\nonumber
&+\frac{16\kappa}{9\pi}\int_{0}^{\infty}dq\ln\left(\frac{q^{2}+pq+p^{2}}{q^{2}-pq+p^{2}}\right)\frac{\ln(q)}{q}t_{+}(q)\\\nonumber
&-\frac{32\kappa}{3\sqrt{3}\pi}\int_{0}^{\infty}dqq(\tilde{C}(q,p,0)+\tilde{V}_{1}(q,p,0)+\tilde{V}_{2}(q,p,0)+\frac{1}{2}\tilde{B}(q,p,0))t_{+}(q)+\cdots\\\nonumber
\end{align}
The function $\tilde{B}(q,p,E)$ is defined as $\tilde{B}(q,p,E)=\frac{1}{8\kappa}B(q,p,E)$, and $\tilde{C}(q,p,E)$, $\tilde{V}_{1}(q,p,E)$, and $\tilde{V}_{2}(q,p,E)$ are defined analogously.  Strictly speaking the integrals over $\tilde{B}(q,p,E)$ and related functions will contain subleading pieces.  However, we only extract numerically those pieces to $\mathcal{O}(\Lambda^{-2})$.  To solve this integral equation we use the ansatz $t_{+}(p)=Cp^{s-1}+A_{+}p^{s-2}+B_{+}\ln(p)p^{s-2}$.  This requires solving~\cite{Bedaque:2002yg}
\begin{equation}
\label{eq:I(s)}
I(s)=\frac{4}{\sqrt{3}\pi}\int_{0}^{\infty}\!\!dx\ln\left(\frac{x^{2}+x+1}{x^{2}-x+1}\right)x^{s-1}=\frac{8}{\sqrt{3}s}\frac{\sin(\frac{\pi s}{6})}{\cos(\frac{\pi s}{2})}.
\end{equation}
For the leading term $Cp^{s-1}$ we find the condition $I(s)=1$.  Solving the resulting transcendental equation for $s$ we find the solutions  $s=\pm is_{0}$, where $s_{0}\simeq 1.0064$ and the constant $C$ is left unsolved in the asymptotic limit since it depends on physics not in the asymptotic regime.  Using the relation $\ln(x)=\frac{\partial}{\partial\alpha}x^{\alpha}\Big{|}_{\alpha=0}$ integrals containing logarithms can be rewritten in the form of Eq.~(\ref{eq:I(s)}), yielding the solution
\begin{align}
\frac{4}{\sqrt{3}\pi}\int_{0}^{\infty}\!\!dx\ln(x)\ln\left(\frac{x^{2}+x+1}{x^{2}-x+1}\right)x^{s-1}&=\frac{\partial}{\partial\alpha}\frac{4}{\sqrt{3}\pi}\int_{0}^{\infty}\!\!dxx^{\alpha}\ln\left(\frac{x^{2}+x+1}{x^{2}-x+1}\right)x^{s-1}\Big{|}_{\alpha=0}\\\nonumber
&=\frac{\partial}{\partial\alpha}I(s+\alpha)\Big{|}_{\alpha=0}=I'(s).
\end{align}
Finally we consider integrals from Coulomb corrections, $\tilde{B}(q,p,0)$, etc.  Integrals over these  functions can be written as an asymptotic series in inverse powers of $p$,
\begin{equation}
\int_{0}^{\infty}dqq^{2}q^{s}\tilde{C}(q,p,0)=\sum_{n=1}^{\infty}\mathcal{J}_{C}(s-n)p^{s-1-n},
\end{equation}
\begin{equation}
\int_{0}^{\infty}dqq^{2}q^{s}\tilde{B}(q,p,0)=\sum_{n=1}^{\infty}\mathcal{J}_{B}(s-n)p^{s-1-n},
\end{equation}
\begin{equation}
\int_{0}^{\infty}dqq^{2}q^{s}\tilde{V}_{1}(q,p,0)=\sum_{n=1}^{\infty}\mathcal{J}_{V_{1}}(s-n)p^{s-1-n},
\end{equation}
\begin{equation}
\int_{0}^{\infty}dqq^{2}q^{s}\tilde{V}_{2}(q,p,0)=\sum_{n=1}^{\infty}\mathcal{J}_{V_{2}}(s-n)p^{s-1-n}.
\end{equation}
The required $\mathcal{O}(\Lambda^{-2})$ contributions are obtained by solving the above integrals for many values of $p$ and then fitting an appropriate polynomial of inverse powers of $p$ to extract the appropriate coefficients. This procedure gives
\begin{align}
&\mathcal{J}_{B}(is_{0}-1)\simeq0.812 - 0.260i\\\nonumber
&\mathcal{J}_{V_{1}}(is_{0}-1)\simeq0.295 - 0.198i\\\nonumber
&\mathcal{J}_{V_{2}}(is_{0}-1)\simeq0.303 - 0.123i\\\nonumber
&\mathcal{J}_{C}(is_{0}-1)\simeq0.186 - 0.113i.
\end{align}
Plugging in the ansatz given above for $t_{+}(p)$ we perform the necessary integrals and the resulting equations give 
\begin{equation}
B_{+}=C\frac{2\kappa}{3}B_{-1}
\end{equation}
and
\begin{equation}
A_{+}=C\left\{\frac{1}{\sqrt{3}}\left(\gamma_{t}+\frac{1}{3}\gamma_{s}+\frac{2}{3}\gamma_{C}\right)B_{-1}+\frac{2\kappa}{3}C_{-1}+\frac{2\kappa}{3}D_{-1}-\frac{16\kappa}{\sqrt{3}\pi}E_{-1}\right\}.
\end{equation}
Note that these coefficients depend upon the constant $C$, which again cannot be solved in the asymptotic limit since it depends on physics not in the asymptotic regime.  The coefficients $B_{-1}$, $C_{-1}$, $D_{-1}$, and $E_{-1}$ are given in terms of the integrals above as
\begin{equation}
B_{-1}=\frac{I(is_{0}-1)}{1-I(is_{0}-1)},
\end{equation}
\begin{equation}
C_{-1}=\frac{I'(is_{0}-1)}{1-I(is_{0}-1)},
\end{equation}
\begin{equation}
D_{-1}=B_{-1}C_{-1},
\end{equation}
and
\begin{equation}
E_{-1}=\frac{\frac{1}{3}\left(2\mathcal{J}_{C}(is_{0}-1)+2\mathcal{J}_{V_{1}}(is_{0}-1)+2\mathcal{J}_{V_{2}}(is_{0}-1)+\mathcal{J}_{B}(is_{0}-1)\right)}{1-I(is_{0}-1)}.
\end{equation}

Collecting all terms up to $\mathcal{O}(\Lambda^{-2})$ we find the following integral equation for $t_{-}(p)$:
\begin{align}
&t_{-}(p)=-\frac{2}{3\pi}\left(\gamma_{t}-\frac{1}{3}\gamma_{s}-\frac{2}{3}\gamma_{C}\right)\frac{1}{p}\int_{0}^{\infty}dq \ln\left(\frac{q^{2}+pq+p^{2}}{q^{2}-pq+p^{2}}\right)Cq^{is_{0}-2}+\\\nonumber
&+\frac{8\kappa}{9\pi}\frac{1}{p}\int_{0}^{\infty}dq\ln(q)\ln\left(\frac{q^{2}+pq+p^{2}}{q^{2}-pq+p^{2}}\right)Cq^{is_{0}-2}\\\nonumber
&-\frac{16\kappa}{3\sqrt{3}\pi}\int_{0}^{\infty}dq(\tilde{C}(q,p,0)+\tilde{V}_{1}(q,p,0)-2\tilde{V}_{2}(q,p,0)+\frac{1}{2}\tilde{B}(q,p,0))Cq^{is_{0}}\\\nonumber
&-\frac{2}{\sqrt{3}\pi}\frac{1}{p}\int_{0}^{\infty}dq\ln\left(\frac{q^{2}+pq+p^{2}}{q^{2}-pq+p^{2}}\right)t_{-}(q) .
\end{align}
Note that the presence of $Cq^{is_{0}-1}$ is merely an insertion of the leading behavior of $t_{+}(q)$.  
Using the ansatz $t_{-}(p)=A_{-}p^{is_{0}-2}+B\ln(p)p^{is_{0}-2}$ we find 
\begin{equation}
B=C\frac{\kappa}{6}\tilde{B}_{-1},
\end{equation}
and
\begin{equation}
A_{-}=C\left\{-\frac{1}{2\sqrt{3}}\left(\gamma_{t}-\frac{1}{3}\gamma_{s}-\frac{2}{3}\gamma_{C}\right)\tilde{B}_{-1}+\frac{\kappa}{6}\tilde{C}_{-1}-\frac{\kappa}{12}\tilde{D}_{-1}-\frac{16\kappa}{\sqrt{3}\pi}\tilde{E}_{-1}\right\}.
\end{equation}
The constants $\tilde{B}_{-1}$, $\tilde{C}_{-1}$, $\tilde{D}_{-1}$, and $\tilde{E}_{-1}$ are again given in terms of the integrals above as
\begin{equation}
\tilde{B}_{-1}=\frac{I(is_{0}-1)}{1+\frac{1}{2}I(is_{0}-1)},
\end{equation}
\begin{equation}
\tilde{C}_{-1}=\frac{I'(is_{0}-1)}{1+\frac{1}{2}I(is_{0}-1)},
\end{equation}
\begin{equation}
\tilde{D}_{-1}=\tilde{B}_{-1}\tilde{C}_{-1},
\end{equation}
and
\begin{equation}
\tilde{E}_{-1}=\frac{\frac{1}{3}\left(\mathcal{J}_{C}(is_{0}-1)+\mathcal{J}_{V_{1}}(is_{0}-1)-2\mathcal{J}_{V_{2}}(is_{0}-1)+\frac{1}{2}\mathcal{J}_{B}(is_{0}-1)\right)}{1+\frac{1}{2}I(is_{0}-1)}.
\end{equation}

Finally we collect all the $\mathcal{O}(\Lambda^{-2})$ terms to find the following integral equation for $t_{\emptyset}(p)$:
\begin{align}
&t_{\emptyset}(p)=-\frac{2}{9\pi}\left(\gamma_{s}-\gamma_{C}\right)\frac{1}{p}\int_{0}^{\infty}dq \ln\left(\frac{q^{2}+pq+p^{2}}{q^{2}-pq+p^{2}}\right)Cq^{is_{0}-2}+\\\nonumber
&+\frac{4\kappa}{9\pi}\frac{1}{p}\int_{0}^{\infty}dq\ln(q)\ln\left(\frac{q^{2}+pq+p^{2}}{q^{2}-pq+p^{2}}\right)Cq^{is_{0}-2}\\\nonumber
&-\frac{8\kappa}{3\sqrt{3}\pi}\int_{0}^{\infty}dq(\tilde{C}(q,p,0)+\tilde{V}_{1}(q,p,0)-2\tilde{V}_{2}(q,p,0)+\frac{1}{2}\tilde{B}(q,p,0))Cq^{is_{0}}\\\nonumber
&-\frac{2}{\sqrt{3}\pi}\frac{1}{p}\int_{0}^{\infty}dq\ln\left(\frac{q^{2}+pq+p^{2}}{q^{2}-pq+p^{2}}\right)t_{\emptyset}(q)
\end{align}

Using the ansatz $t_{\emptyset}(p)=A_{\emptyset}p^{is_{0}-2}+B\ln(p)p^{is_{0}-2}$ we find that $B$ is the same as in the $t_{-}(p)$ ansatz, and $A_{\emptyset}$ is given by
\begin{equation}
A_{\emptyset}=C\left\{-\frac{1}{6\sqrt{3}}\left(\gamma_{s}-\gamma_{C}\right)\tilde{B}_{-1}+\frac{\kappa}{6}\tilde{C}_{-1}-\frac{\kappa}{12}\tilde{D}_{-1}-\frac{8\kappa}{\sqrt{3}\pi}\tilde{E}_{-1}\right\}.
\end{equation}


\begin{thebibliography}{52}
\expandafter\ifx\csname natexlab\endcsname\relax\def\natexlab#1{#1}\fi
\expandafter\ifx\csname bibnamefont\endcsname\relax
  \def\bibnamefont#1{#1}\fi
\expandafter\ifx\csname bibfnamefont\endcsname\relax
  \def\bibfnamefont#1{#1}\fi
\expandafter\ifx\csname citenamefont\endcsname\relax
  \def\citenamefont#1{#1}\fi
\expandafter\ifx\csname url\endcsname\relax
  \def\url#1{\texttt{#1}}\fi
\expandafter\ifx\csname urlprefix\endcsname\relax\def\urlprefix{URL }\fi
\providecommand{\bibinfo}[2]{#2}
\providecommand{\eprint}[2][]{\url{#2}}

\bibitem[{\citenamefont{Huttel et~al.}(1983)\citenamefont{Huttel, Arnold, Berg,
  Krause, Ulbricht, and Clausnitzer}}]{Huttel:1983}
\bibinfo{author}{\bibfnamefont{E.}~\bibnamefont{Huttel}},
  \bibinfo{author}{\bibfnamefont{W.}~\bibnamefont{Arnold}},
  \bibinfo{author}{\bibfnamefont{H.}~\bibnamefont{Berg}},
  \bibinfo{author}{\bibfnamefont{H.}~\bibnamefont{Krause}},
  \bibinfo{author}{\bibfnamefont{J.}~\bibnamefont{Ulbricht}}, \bibnamefont{and}
  \bibinfo{author}{\bibfnamefont{G.}~\bibnamefont{Clausnitzer}},
  \bibinfo{journal}{Nucl. Phys. A} \textbf{\bibinfo{volume}{406}},
  \bibinfo{pages}{435} (\bibinfo{year}{1983}).

\bibitem[{\citenamefont{McAninch et~al.}(1994)\citenamefont{McAninch, Lamm, and
  Haeberli}}]{PhysRevC.50.589}
\bibinfo{author}{\bibfnamefont{J.~E.} \bibnamefont{McAninch}},
  \bibinfo{author}{\bibfnamefont{L.~O.} \bibnamefont{Lamm}}, \bibnamefont{and}
  \bibinfo{author}{\bibfnamefont{W.}~\bibnamefont{Haeberli}},
  \bibinfo{journal}{Phys. Rev. C} \textbf{\bibinfo{volume}{50}},
  \bibinfo{pages}{589} (\bibinfo{year}{1994}),
  \urlprefix\url{http://link.aps.org/doi/10.1103/PhysRevC.50.589}.

\bibitem[{\citenamefont{Shimizu et~al.}(1995)\citenamefont{Shimizu, Sagara,
  Nakamura, Maeda, Miwa, Nishimori, Ueno, Nakashima, and
  Morinobu}}]{PhysRevC.52.1193}
\bibinfo{author}{\bibfnamefont{S.}~\bibnamefont{Shimizu}},
  \bibinfo{author}{\bibfnamefont{K.}~\bibnamefont{Sagara}},
  \bibinfo{author}{\bibfnamefont{H.}~\bibnamefont{Nakamura}},
  \bibinfo{author}{\bibfnamefont{K.}~\bibnamefont{Maeda}},
  \bibinfo{author}{\bibfnamefont{T.}~\bibnamefont{Miwa}},
  \bibinfo{author}{\bibfnamefont{N.}~\bibnamefont{Nishimori}},
  \bibinfo{author}{\bibfnamefont{S.}~\bibnamefont{Ueno}},
  \bibinfo{author}{\bibfnamefont{T.}~\bibnamefont{Nakashima}},
  \bibnamefont{and} \bibinfo{author}{\bibfnamefont{S.}~\bibnamefont{Morinobu}},
  \bibinfo{journal}{Phys. Rev. C} \textbf{\bibinfo{volume}{52}},
  \bibinfo{pages}{1193} (\bibinfo{year}{1995}),
  \urlprefix\url{http://link.aps.org/doi/10.1103/PhysRevC.52.1193}.

\bibitem[{\citenamefont{Kievsky et~al.}(1996)\citenamefont{Kievsky, Rosati,
  Tornow, and Viviani}}]{Kievsky:1996ca}
\bibinfo{author}{\bibfnamefont{A.}~\bibnamefont{Kievsky}},
  \bibinfo{author}{\bibfnamefont{S.}~\bibnamefont{Rosati}},
  \bibinfo{author}{\bibfnamefont{W.}~\bibnamefont{Tornow}}, \bibnamefont{and}
  \bibinfo{author}{\bibfnamefont{M.}~\bibnamefont{Viviani}},
  \bibinfo{journal}{Nucl. Phys. A} \textbf{\bibinfo{volume}{607}},
  \bibinfo{pages}{402} (\bibinfo{year}{1996}).

\bibitem[{\citenamefont{Huber and Friar}(1998)}]{Huber:1998hu}
\bibinfo{author}{\bibfnamefont{D.}~\bibnamefont{Huber}} \bibnamefont{and}
  \bibinfo{author}{\bibfnamefont{J.~L.} \bibnamefont{Friar}},
  \bibinfo{journal}{Phys. Rev. C} \textbf{\bibinfo{volume}{58}},
  \bibinfo{pages}{674} (\bibinfo{year}{1998}), \eprint{nucl-th/9803038}.

\bibitem[{\citenamefont{Entem et~al.}(2002)\citenamefont{Entem, Machleidt, and
  Wita{\l}a}}]{Entem:2001tj}
\bibinfo{author}{\bibfnamefont{D.}~\bibnamefont{Entem}},
  \bibinfo{author}{\bibfnamefont{R.}~\bibnamefont{Machleidt}},
  \bibnamefont{and} \bibinfo{author}{\bibfnamefont{H.}~\bibnamefont{Wita{\l}a}},
  \bibinfo{journal}{Phys. Rev. C} \textbf{\bibinfo{volume}{65}},
  \bibinfo{pages}{064005} (\bibinfo{year}{2002}), \eprint{nucl-th/0111033}.

\bibitem[{\citenamefont{Gl{\"o}ckle et~al.}(1996)\citenamefont{Gl{\"o}ckle,
  Wita{\l}a, Huber, Kamada, and Golak}}]{Gloeckle1996107}
\bibinfo{author}{\bibfnamefont{W.}~\bibnamefont{Gl{\"o}ckle}},
  \bibinfo{author}{\bibfnamefont{H.}~\bibnamefont{Wita{\l}a}},
  \bibinfo{author}{\bibfnamefont{D.}~\bibnamefont{Huber}},
  \bibinfo{author}{\bibfnamefont{H.}~\bibnamefont{Kamada}}, \bibnamefont{and}
  \bibinfo{author}{\bibfnamefont{J.}~\bibnamefont{Golak}},
  \bibinfo{journal}{Physics Reports} \textbf{\bibinfo{volume}{274}},
  \bibinfo{pages}{107 } (\bibinfo{year}{1996}), ISSN \bibinfo{issn}{0370-1573},
  \urlprefix\url{http://www.sciencedirect.com/science/article/pii/037015739500%
0852}.

\bibitem[{\citenamefont{van Kolck}(1998)}]{vanKolck:1997ut}
\bibinfo{author}{\bibfnamefont{U.}~\bibnamefont{van Kolck}}, in
  \emph{\bibinfo{booktitle}{Chiral Dynamics: Theory and Experiment}}, edited by
  \bibinfo{editor}{\bibfnamefont{A.}~\bibnamefont{Bernstein}},
  \bibinfo{editor}{\bibfnamefont{D.}~\bibnamefont{Drechsel}}, \bibnamefont{and}
  \bibinfo{editor}{\bibfnamefont{T.}~\bibnamefont{Walcher}}
  (\bibinfo{publisher}{Springer Berlin Heidelberg}, \bibinfo{year}{1998}), vol.
  \bibinfo{volume}{513} of \emph{\bibinfo{series}{Lecture Notes in Physics}},
  pp. \bibinfo{pages}{62--77}, ISBN \bibinfo{isbn}{978-3-540-64716-4},
  \eprint{hep-ph/9711222},
  \urlprefix\url{http://dx.doi.org/10.1007/BFb0104898}.

\bibitem[{\citenamefont{van Kolck}(1999)}]{vanKolck:1998bw}
\bibinfo{author}{\bibfnamefont{U.}~\bibnamefont{van Kolck}},
  \bibinfo{journal}{Nucl.Phys.} \textbf{\bibinfo{volume}{A645}},
  \bibinfo{pages}{273} (\bibinfo{year}{1999}), \eprint{nucl-th/9808007}.

\bibitem[{\citenamefont{Gegelia}(1998)}]{Gegelia:1998gn}
\bibinfo{author}{\bibfnamefont{J.}~\bibnamefont{Gegelia}},
  \bibinfo{journal}{Phys. Lett. B} \textbf{\bibinfo{volume}{429}},
  \bibinfo{pages}{227} (\bibinfo{year}{1998}).

\bibitem[{\citenamefont{Kaplan et~al.}(1998)\citenamefont{Kaplan, Savage, and
  Wise}}]{Kaplan:1998tg}
\bibinfo{author}{\bibfnamefont{D.~B.} \bibnamefont{Kaplan}},
  \bibinfo{author}{\bibfnamefont{M.~J.} \bibnamefont{Savage}},
  \bibnamefont{and} \bibinfo{author}{\bibfnamefont{M.~B.} \bibnamefont{Wise}},
  \bibinfo{journal}{Phys. Lett. B} \textbf{\bibinfo{volume}{424}},
  \bibinfo{pages}{390} (\bibinfo{year}{1998}), \eprint{nucl-th/9801034}.

\bibitem[{\citenamefont{Mehen and Stewart}(1999)}]{Mehen:1998tp}
\bibinfo{author}{\bibfnamefont{T.}~\bibnamefont{Mehen}} \bibnamefont{and}
  \bibinfo{author}{\bibfnamefont{I.~W.} \bibnamefont{Stewart}},
  \bibinfo{journal}{Phys. Rev. C} \textbf{\bibinfo{volume}{59}},
  \bibinfo{pages}{2365} (\bibinfo{year}{1999}), \eprint{nucl-th/9809095}.

\bibitem[{\citenamefont{Kong and Ravndal}(2000)}]{Kong:1999sf}
\bibinfo{author}{\bibfnamefont{X.-w.} \bibnamefont{Kong}} \bibnamefont{and}
  \bibinfo{author}{\bibfnamefont{F.}~\bibnamefont{Ravndal}},
  \bibinfo{journal}{Nucl. Phys. A} \textbf{\bibinfo{volume}{665}},
  \bibinfo{pages}{137} (\bibinfo{year}{2000}), \eprint{hep-ph/9903523}.

\bibitem[{\citenamefont{Chen et~al.}(1999)\citenamefont{Chen, Rupak, and
  Savage}}]{Chen:1999tn}
\bibinfo{author}{\bibfnamefont{J.-W.} \bibnamefont{Chen}},
  \bibinfo{author}{\bibfnamefont{G.}~\bibnamefont{Rupak}}, \bibnamefont{and}
  \bibinfo{author}{\bibfnamefont{M.~J.} \bibnamefont{Savage}},
  \bibinfo{journal}{Nucl. Phys. A} \textbf{\bibinfo{volume}{653}},
  \bibinfo{pages}{386} (\bibinfo{year}{1999}), \eprint{nucl-th/9902056}.

\bibitem[{\citenamefont{Ando and Hyun}(2005)}]{Ando:2004mm}
\bibinfo{author}{\bibfnamefont{S.-i.} \bibnamefont{Ando}} \bibnamefont{and}
  \bibinfo{author}{\bibfnamefont{C.~H.} \bibnamefont{Hyun}},
  \bibinfo{journal}{Phys. Rev. C} \textbf{\bibinfo{volume}{72}},
  \bibinfo{pages}{014008} (\bibinfo{year}{2005}), \eprint{nucl-th/0407103}.

\bibitem[{\citenamefont{Ando et~al.}(2007)\citenamefont{Ando, Shin, Hyun, and
  Hong}}]{Ando:2007fh}
\bibinfo{author}{\bibfnamefont{S.-i.} \bibnamefont{Ando}},
  \bibinfo{author}{\bibfnamefont{J.~W.} \bibnamefont{Shin}},
  \bibinfo{author}{\bibfnamefont{C.~H.} \bibnamefont{Hyun}}, \bibnamefont{and}
  \bibinfo{author}{\bibfnamefont{S.~W.} \bibnamefont{Hong}},
  \bibinfo{journal}{Phys. Rev. C} \textbf{\bibinfo{volume}{76}},
  \bibinfo{pages}{064001} (\bibinfo{year}{2007}), \eprint{0704.2312}.

\bibitem[{\citenamefont{Chen and Savage}(1999)}]{Chen:1999bg}
\bibinfo{author}{\bibfnamefont{J.-W.} \bibnamefont{Chen}} \bibnamefont{and}
  \bibinfo{author}{\bibfnamefont{M.~J.} \bibnamefont{Savage}},
  \bibinfo{journal}{Phys. Rev. C} \textbf{\bibinfo{volume}{60}},
  \bibinfo{pages}{065205} (\bibinfo{year}{1999}), \eprint{nucl-th/9907042}.

\bibitem[{\citenamefont{Rupak}(2000)}]{Rupak:1999rk}
\bibinfo{author}{\bibfnamefont{G.}~\bibnamefont{Rupak}},
  \bibinfo{journal}{Nucl. Phys. A} \textbf{\bibinfo{volume}{678}},
  \bibinfo{pages}{405} (\bibinfo{year}{2000}), \eprint{nucl-th/9911018}.

\bibitem[{\citenamefont{Ando et~al.}(2006)\citenamefont{Ando, Cyburt, Hong, and
  Hyun}}]{Ando:2005cz}
\bibinfo{author}{\bibfnamefont{S.}~\bibnamefont{Ando}},
  \bibinfo{author}{\bibfnamefont{R.}~\bibnamefont{Cyburt}},
  \bibinfo{author}{\bibfnamefont{S.}~\bibnamefont{Hong}}, \bibnamefont{and}
  \bibinfo{author}{\bibfnamefont{C.}~\bibnamefont{Hyun}},
  \bibinfo{journal}{Phys. Rev. C} \textbf{\bibinfo{volume}{74}},
  \bibinfo{pages}{025809} (\bibinfo{year}{2006}), \eprint{nucl-th/0511074}.

\bibitem[{\citenamefont{Savage}(2001)}]{Savage:2000iv}
\bibinfo{author}{\bibfnamefont{M.~J.} \bibnamefont{Savage}},
  \bibinfo{journal}{Nucl.Phys.} \textbf{\bibinfo{volume}{A695}},
  \bibinfo{pages}{365} (\bibinfo{year}{2001}), \eprint{nucl-th/0012043}.

\bibitem[{\citenamefont{Phillips et~al.}(2009)\citenamefont{Phillips,
  Schindler, and Springer}}]{Phillips:2008hn}
\bibinfo{author}{\bibfnamefont{D.~R.} \bibnamefont{Phillips}},
  \bibinfo{author}{\bibfnamefont{M.~R.} \bibnamefont{Schindler}},
  \bibnamefont{and} \bibinfo{author}{\bibfnamefont{R.~P.}
  \bibnamefont{Springer}}, \bibinfo{journal}{Nucl. Phys. A}
  \textbf{\bibinfo{volume}{822}}, \bibinfo{pages}{1} (\bibinfo{year}{2009}),
  \eprint{0812.2073}.

\bibitem[{\citenamefont{Schindler and Springer}(2010)}]{Schindler:2009wd}
\bibinfo{author}{\bibfnamefont{M.~R.} \bibnamefont{Schindler}}
  \bibnamefont{and} \bibinfo{author}{\bibfnamefont{R.~P.}
  \bibnamefont{Springer}}, \bibinfo{journal}{Nucl. Phys. A}
  \textbf{\bibinfo{volume}{846}}, \bibinfo{pages}{51} (\bibinfo{year}{2010}),
  \eprint{0907.5358}.

\bibitem[{\citenamefont{Shin et~al.}(2010)\citenamefont{Shin, Ando, and
  Hyun}}]{Shin:2009hi}
\bibinfo{author}{\bibfnamefont{J.}~\bibnamefont{Shin}},
  \bibinfo{author}{\bibfnamefont{S.}~\bibnamefont{Ando}}, \bibnamefont{and}
  \bibinfo{author}{\bibfnamefont{C.}~\bibnamefont{Hyun}},
  \bibinfo{journal}{Phys. Rev. C} \textbf{\bibinfo{volume}{81}},
  \bibinfo{pages}{055501} (\bibinfo{year}{2010}), \eprint{0907.3995}.

\bibitem[{\citenamefont{Kong and Ravndal}(2001)}]{Kong:2000px}
\bibinfo{author}{\bibfnamefont{X.-w.} \bibnamefont{Kong}} \bibnamefont{and}
  \bibinfo{author}{\bibfnamefont{F.}~\bibnamefont{Ravndal}},
  \bibinfo{journal}{Phys. Rev. C} \textbf{\bibinfo{volume}{64}},
  \bibinfo{pages}{044002} (\bibinfo{year}{2001}), \eprint{nucl-th/0004038}.

\bibitem[{\citenamefont{Butler et~al.}(2001)\citenamefont{Butler, Chen, and
  Kong}}]{Butler:2000zp}
\bibinfo{author}{\bibfnamefont{M.}~\bibnamefont{Butler}},
  \bibinfo{author}{\bibfnamefont{J.-W.} \bibnamefont{Chen}}, \bibnamefont{and}
  \bibinfo{author}{\bibfnamefont{X.-w.} \bibnamefont{Kong}},
  \bibinfo{journal}{Phys. Rev. C} \textbf{\bibinfo{volume}{63}},
  \bibinfo{pages}{035501} (\bibinfo{year}{2001}), \eprint{nucl-th/0008032}.

\bibitem[{\citenamefont{Ando et~al.}(2008)\citenamefont{Ando, Shin, Hyun, Hong,
  and Kubodera}}]{Ando:2008va}
\bibinfo{author}{\bibfnamefont{S.}~\bibnamefont{Ando}},
  \bibinfo{author}{\bibfnamefont{J.}~\bibnamefont{Shin}},
  \bibinfo{author}{\bibfnamefont{C.}~\bibnamefont{Hyun}},
  \bibinfo{author}{\bibfnamefont{S.}~\bibnamefont{Hong}}, \bibnamefont{and}
  \bibinfo{author}{\bibfnamefont{K.}~\bibnamefont{Kubodera}},
  \bibinfo{journal}{Phys. Lett. B} \textbf{\bibinfo{volume}{668}},
  \bibinfo{pages}{187} (\bibinfo{year}{2008}), \eprint{0801.4330}.

\bibitem[{\citenamefont{Chen et~al.}(2013)\citenamefont{Chen, Liu, and
  Yu}}]{Chen:2012hm}
\bibinfo{author}{\bibfnamefont{J.-W.} \bibnamefont{Chen}},
  \bibinfo{author}{\bibfnamefont{C.-P.} \bibnamefont{Liu}}, \bibnamefont{and}
  \bibinfo{author}{\bibfnamefont{S.-H.} \bibnamefont{Yu}},
  \bibinfo{journal}{Phys. Lett. B} \textbf{\bibinfo{volume}{720}},
  \bibinfo{pages}{385} (\bibinfo{year}{2013}), \eprint{1209.2552}.

\bibitem[{\citenamefont{Bedaque and Grie{\ss}hammer}(2000)}]{Bedaque:1999vb}
\bibinfo{author}{\bibfnamefont{P.~F.} \bibnamefont{Bedaque}} \bibnamefont{and}
  \bibinfo{author}{\bibfnamefont{H.~W.} \bibnamefont{Grie{\ss}hammer}},
  \bibinfo{journal}{Nucl. Phys. A} \textbf{\bibinfo{volume}{671}},
  \bibinfo{pages}{357} (\bibinfo{year}{2000}), \eprint{nucl-th/9907077}.

\bibitem[{\citenamefont{Gabbiani et~al.}(2000)\citenamefont{Gabbiani, Bedaque,
  and Grie{\ss}hammer}}]{Gabbiani:1999yv}
\bibinfo{author}{\bibfnamefont{F.}~\bibnamefont{Gabbiani}},
  \bibinfo{author}{\bibfnamefont{P.~F.} \bibnamefont{Bedaque}},
  \bibnamefont{and} \bibinfo{author}{\bibfnamefont{H.~W.}
  \bibnamefont{Grie{\ss}hammer}}, \bibinfo{journal}{Nucl. Phys. A}
  \textbf{\bibinfo{volume}{675}}, \bibinfo{pages}{601} (\bibinfo{year}{2000}),
  \eprint{nucl-th/9911034}.

\bibitem[{\citenamefont{Rupak and Kong}(2003)}]{Rupak:2001ci}
\bibinfo{author}{\bibfnamefont{G.}~\bibnamefont{Rupak}} \bibnamefont{and}
  \bibinfo{author}{\bibfnamefont{X.-w.} \bibnamefont{Kong}},
  \bibinfo{journal}{Nucl. Phys. A} \textbf{\bibinfo{volume}{717}},
  \bibinfo{pages}{73} (\bibinfo{year}{2003}), \eprint{nucl-th/0108059}.

\bibitem[{\citenamefont{Bedaque et~al.}(2003)\citenamefont{Bedaque, Rupak,
  Grie{\ss}hammer, and Hammer}}]{Bedaque:2002yg}
\bibinfo{author}{\bibfnamefont{P.~F.} \bibnamefont{Bedaque}},
  \bibinfo{author}{\bibfnamefont{G.}~\bibnamefont{Rupak}},
  \bibinfo{author}{\bibfnamefont{H.~W.} \bibnamefont{Grie{\ss}hammer}},
  \bibnamefont{and} \bibinfo{author}{\bibfnamefont{H.-W.}
  \bibnamefont{Hammer}}, \bibinfo{journal}{Nucl. Phys. A}
  \textbf{\bibinfo{volume}{714}}, \bibinfo{pages}{589} (\bibinfo{year}{2003}),
  \eprint{nucl-th/0207034}.

\bibitem[{\citenamefont{Grie{\ss}hammer}(2004)}]{Griesshammer:2004pe}
\bibinfo{author}{\bibfnamefont{H.~W.} \bibnamefont{Grie{\ss}hammer}},
  \bibinfo{journal}{Nucl. Phys. A} \textbf{\bibinfo{volume}{744}},
  \bibinfo{pages}{192} (\bibinfo{year}{2004}), \eprint{nucl-th/0404073}.

\bibitem[{\citenamefont{K{\"o}nig and Hammer}(2011)}]{Konig:2011yq}
\bibinfo{author}{\bibfnamefont{S.}~\bibnamefont{K{\"o}nig}} \bibnamefont{and}
  \bibinfo{author}{\bibfnamefont{H.-W.} \bibnamefont{Hammer}},
  \bibinfo{journal}{Phys. Rev. C} \textbf{\bibinfo{volume}{83}},
  \bibinfo{pages}{064001} (\bibinfo{year}{2011}), \eprint{1101.5939}.

\bibitem[{\citenamefont{Vanasse}(2013)}]{Vanasse:2013sda}
\bibinfo{author}{\bibfnamefont{J.}~\bibnamefont{Vanasse}},
  \bibinfo{journal}{Phys. Rev. C} \textbf{\bibinfo{volume}{88}},
  \bibinfo{pages}{044001} (\bibinfo{year}{2013}), \eprint{1305.0283}.

\bibitem[{\citenamefont{Ando and Birse}(2010)}]{Ando:2010wq}
\bibinfo{author}{\bibfnamefont{S.-i.} \bibnamefont{Ando}} \bibnamefont{and}
  \bibinfo{author}{\bibfnamefont{M.~C.} \bibnamefont{Birse}},
  \bibinfo{journal}{J. Phys. G} \textbf{\bibinfo{volume}{37}},
  \bibinfo{pages}{105108} (\bibinfo{year}{2010}), \eprint{1003.4383}.

\bibitem[{\citenamefont{Grie{\ss}hammer
  et~al.}(2012)\citenamefont{Grie{\ss}hammer, Schindler, and
  Springer}}]{Griesshammer:2011md}
\bibinfo{author}{\bibfnamefont{H.~W.} \bibnamefont{Grie{\ss}hammer}},
  \bibinfo{author}{\bibfnamefont{M.~R.} \bibnamefont{Schindler}},
  \bibnamefont{and} \bibinfo{author}{\bibfnamefont{R.~P.}
  \bibnamefont{Springer}}, \bibinfo{journal}{Eur. Phys. J. A}
  \textbf{\bibinfo{volume}{48}}, \bibinfo{pages}{7} (\bibinfo{year}{2012}),
  \eprint{1109.5667}.

\bibitem[{\citenamefont{Vanasse}(2012)}]{Vanasse:2011nd}
\bibinfo{author}{\bibfnamefont{J.}~\bibnamefont{Vanasse}},
  \bibinfo{journal}{Phys. Rev. C} \textbf{\bibinfo{volume}{86}},
  \bibinfo{pages}{014001} (\bibinfo{year}{2012}), \eprint{1110.1039}.

\bibitem[{\citenamefont{K{\"o}nig}(2013)}]{Koenig:2013}
\bibinfo{author}{\bibfnamefont{S.}~\bibnamefont{K{\"o}nig}}, Ph.D. thesis,
  \bibinfo{school}{Universit{\"a}t Bonn} (\bibinfo{year}{2013}).

  
\bibitem[{\citenamefont{K{\"o}nig et~al.}(2014)\citenamefont{K{\"o}nig,
  Grie{\ss}hammer, and Hammer}}]{Konig:2014ufa}
\bibinfo{author}{\bibfnamefont{S.}~\bibnamefont{K{\"o}nig}},
  \bibinfo{author}{\bibfnamefont{H.~W.} \bibnamefont{Grie{\ss}hammer}},
  \bibnamefont{and} \bibinfo{author}{\bibfnamefont{H.-W.} \bibnamefont{Hammer}}, \eprint{1405.7961}.

\bibitem[{\citenamefont{Phillips et~al.}(2000)\citenamefont{Phillips, Rupak,
  and Savage}}]{Phillips:1999hh}
\bibinfo{author}{\bibfnamefont{D.~R.} \bibnamefont{Phillips}},
  \bibinfo{author}{\bibfnamefont{G.}~\bibnamefont{Rupak}}, \bibnamefont{and}
  \bibinfo{author}{\bibfnamefont{M.~J.} \bibnamefont{Savage}},
  \bibinfo{journal}{Phys. Lett. B} \textbf{\bibinfo{volume}{473}},
  \bibinfo{pages}{209} (\bibinfo{year}{2000}), \eprint{nucl-th/9908054}.

\bibitem[{\citenamefont{Holstein}(2009)}]{Holstein:2009zzb}
\bibinfo{author}{\bibfnamefont{B.}~\bibnamefont{Holstein}},
  \bibinfo{journal}{Eur. Phys. J. A} \textbf{\bibinfo{volume}{41}},
  \bibinfo{pages}{279} (\bibinfo{year}{2009}).

\bibitem[{\citenamefont{Hoferichter}(2010)}]{Hoferichter-BoxDiag:2010}
\bibinfo{author}{\bibfnamefont{M.}~\bibnamefont{Hoferichter}}
  (\bibinfo{year}{2010}), \bibinfo{note}{private communication}.

\bibitem[{\citenamefont{Ji and Phillips}(2013)}]{Ji:2012nj}
\bibinfo{author}{\bibfnamefont{C.}~\bibnamefont{Ji}} \bibnamefont{and}
  \bibinfo{author}{\bibfnamefont{D.~R.} \bibnamefont{Phillips}},
  \bibinfo{journal}{Few-Body Syst.} \textbf{\bibinfo{volume}{54}},
  \bibinfo{pages}{2317} (\bibinfo{year}{2013}), \eprint{1212.1845}.

\bibitem[{\citenamefont{Amado}(1966)}]{PhysRev.141.902}
\bibinfo{author}{\bibfnamefont{R.~D.} \bibnamefont{Amado}},
  \bibinfo{journal}{Phys. Rev.} \textbf{\bibinfo{volume}{141}},
  \bibinfo{pages}{902} (\bibinfo{year}{1966}),
  \urlprefix\url{http://link.aps.org/doi/10.1103/PhysRev.141.902}.

\bibitem[{\citenamefont{Ji et~al.}(2012)\citenamefont{Ji, Phillips, and
  Platter}}]{Ji:2011qg}
\bibinfo{author}{\bibfnamefont{C.}~\bibnamefont{Ji}},
  \bibinfo{author}{\bibfnamefont{D.~R.} \bibnamefont{Phillips}},
  \bibnamefont{and} \bibinfo{author}{\bibfnamefont{L.}~\bibnamefont{Platter}},
  \bibinfo{journal}{Ann. Phys. (N.Y.)} \textbf{\bibinfo{volume}{327}},
  \bibinfo{pages}{1803} (\bibinfo{year}{2012}), \eprint{1106.3837}.

\bibitem[{\citenamefont{Bedaque et~al.}(2000)\citenamefont{Bedaque, Hammer, and
  van Kolck}}]{Bedaque:1999ve}
\bibinfo{author}{\bibfnamefont{P.~F.} \bibnamefont{Bedaque}},
  \bibinfo{author}{\bibfnamefont{H.-W.} \bibnamefont{Hammer}},
  \bibnamefont{and} \bibinfo{author}{\bibfnamefont{U.}~\bibnamefont{van
  Kolck}}, \bibinfo{journal}{Nucl. Phys. A} \textbf{\bibinfo{volume}{676}},
  \bibinfo{pages}{357} (\bibinfo{year}{2000}), \eprint{nucl-th/9906032}.

\bibitem[{\citenamefont{Wigner}(1937)}]{PhysRev.51.106}
\bibinfo{author}{\bibfnamefont{E.}~\bibnamefont{Wigner}},
  \bibinfo{journal}{Phys. Rev.} \textbf{\bibinfo{volume}{51}},
  \bibinfo{pages}{106} (\bibinfo{year}{1937}),
  \urlprefix\url{http://link.aps.org/doi/10.1103/PhysRev.51.106}.

\bibitem[{\citenamefont{Mehen et~al.}(1999)\citenamefont{Mehen, Stewart, and
  Wise}}]{Mehen:1999qs}
\bibinfo{author}{\bibfnamefont{T.}~\bibnamefont{Mehen}},
  \bibinfo{author}{\bibfnamefont{I.~W.} \bibnamefont{Stewart}},
  \bibnamefont{and} \bibinfo{author}{\bibfnamefont{M.~B.} \bibnamefont{Wise}},
  \bibinfo{journal}{Phys.Rev.Lett.} \textbf{\bibinfo{volume}{83}},
  \bibinfo{pages}{931} (\bibinfo{year}{1999}), \eprint{hep-ph/9902370}.

\bibitem[{\citenamefont{Bedaque et~al.}(1999)\citenamefont{Bedaque, Hammer, and
  van Kolck}}]{Bedaque:1998km}
\bibinfo{author}{\bibfnamefont{P.~F.} \bibnamefont{Bedaque}},
  \bibinfo{author}{\bibfnamefont{H.-W.} \bibnamefont{Hammer}},
  \bibnamefont{and} \bibinfo{author}{\bibfnamefont{U.}~\bibnamefont{van
  Kolck}}, \bibinfo{journal}{Nucl. Phys. A} \textbf{\bibinfo{volume}{646}},
  \bibinfo{pages}{444} (\bibinfo{year}{1999}), \eprint{nucl-th/9811046}.

\bibitem[{\citenamefont{Braaten et~al.}(2011)\citenamefont{Braaten, Kang, and
  Platter}}]{Braaten:2011sz}
\bibinfo{author}{\bibfnamefont{E.}~\bibnamefont{Braaten}},
  \bibinfo{author}{\bibfnamefont{D.}~\bibnamefont{Kang}}, \bibnamefont{and}
  \bibinfo{author}{\bibfnamefont{L.}~\bibnamefont{Platter}},
  \bibinfo{journal}{Phys. Rev. Lett.} \textbf{\bibinfo{volume}{106}},
  \bibinfo{pages}{153005} (\bibinfo{year}{2011}), \eprint{1101.2854}.

\bibitem[{\citenamefont{Hammer and Mehen}(2001)}]{Hammer:2001gh}
\bibinfo{author}{\bibfnamefont{H.-W.} \bibnamefont{Hammer}} \bibnamefont{and}
  \bibinfo{author}{\bibfnamefont{T.}~\bibnamefont{Mehen}},
  \bibinfo{journal}{Phys. Lett. B} \textbf{\bibinfo{volume}{516}},
  \bibinfo{pages}{353} (\bibinfo{year}{2001}), \eprint{nucl-th/0105072}.

\bibitem[{\citenamefont{Arvieux}(1974)}]{Arvieux:1974}
\bibinfo{author}{\bibfnamefont{J.}~\bibnamefont{Arvieux}},
  \bibinfo{journal}{Nucl. Phys. A} \textbf{\bibinfo{volume}{221}},
  \bibinfo{pages}{253} (\bibinfo{year}{1974}).

\end{thebibliography}

\end{document}